\documentclass[onecolumn]{pasj01}
\Received{$\langle$reception date$\rangle$}
\Accepted{$\langle$acception date$\rangle$}
\Published{$\langle$publication date$\rangle$}
\usepackage{lineno}

\begin{document}

\title{Searching for the Signature of Fast Radio Burst by \textit{Swift}/XRT X-ray Afterglow Light Curve }
\author{Hsien-chieh \textsc{Shen},$^{*}$
        Takanori \textsc{Sakamoto},$^{*}$
        Motoko \textsc{Serino},$^{*}$ and
        Yuri \textsc{Sato}}%
\altaffiltext{}{Department of Physical Sciences, Aoyama Gakuin University, 5-10-1 Fuchinobe, Chuo-ku,
Sagamihara, Kanagawa 252-5258, Japan }
\email{ksei@phys.aoyama.ac.jp, tsakamoto@phys.aoyama.ac.jp, mserino@phys.aoyama.ac.jp}


\maketitle

\begin{abstract}
A new type of cosmological transient, dubbed fast radio bursts (FRBs), was recently discovered. The source of FRBs is still unknown. One possible scenario of an FRB is the collapse of a spinning supra-massive neutron star. \citet{BingZhang2014} suggests that the collapse can happen shortly (hundreds to thousands of seconds) after the birth of supra-massive neutron stars. The signatures can be visible in X-ray afterglows of long and short gamma-ray bursts (GRBs). For instance, a sudden drop (decay index steeper than $-3$ to $-9$) from a shallow decay (decay index shallower than $-1$) in the X-ray afterglow flux can indicate the event. We selected the X-ray afterglow light curves with a steep decay after the shallow decay phase from the Swift/XRT GRB catalog. We analyzed when the decay index changed suddenly by fitting these light curves to double power-law functions and compared it with the onset of FRBs. We found none of our GRB samples match the onset of FRBs. 
\end{abstract}
\section{ Introduction }
Fast radio bursts (FRBs) are enigmatic transient radio bursts with a typical frequency of 100 MHz to GHz. These radio bursts have a typical duration of several milliseconds, a high dispersion measure (DM), and the inferred total energy release of ${\sim}$ ${10^{39}}$ erg. According to the observational properties, FRB's origin is considered a compact object located at a cosmological distance. Since the discovery of Lorimer burst in 2006 \citep{Lorimer2007}, hundreds of FRBs have been identified by radio observatories all over the world, such as the Canadian Hydrogen Intensity Mapping Experiment (CHIME) \citep{CHIMEFRB_Collaboration} and Australian Square Kilometre Array Pathfinder (ASKAP) \citep{ASKAP}. However, the origin and the emission mechanism are still under debate. There are more than 50 theoretical models for FRBs, and young magnetars have been put forward as the leading source candidate for repeating FRBs \citep{Michilli2018}. In contrast, the origin of non-repeating FRBs is still unknown.

\citet{Moroianu2023} show that a binary neutron star (BNS) merger could be one of the coincidence sources of non-repeating FRBs. A non-repeating CHIME/FRB event, FRB~20190425A, located within the gravitational waves's (GW) sky localization area of LIGO, was detected 2.5 hours after the GW event, GW~190425. According to the GW data, GW~190425 is consistent with a BNS. However, the chirp mass and the total mass are significantly larger than those of any known BNS system  \citep{Abbott2020}. FRB 20190425A is a bright FRB with a fluence of 31.6 $\pm$ 4.2 Jy ms and a duration of 380 $\pm$ 2 $\mu$s, and has an unusually low DM of 128.2 pc cm$^{-3}$. Considering the temporal, spatial, and DM of GW~190425 and FRB~20190425A, \citet{Moroianu2023} claims the chance coincidence between unrelated FRB and GW events to be 0.0052 (2.8 $\sigma$). 


Although no FRB-associated GRB event has been reported, \citet{Rowlinson2023} claimed the detection of a coherent radio flash 76.6 minutes after a short GRB event, GRB 201006A. This radio flash is offset by 27$^{\prime\prime}$ from the GRB location, which has a chance probability of ${\sim}$ 0.5 $\%$ (2.6 $\sigma$), considering measurement uncertainties. However, its low significance detection warns against a further multi-wavelength search to claim the association between an FRB and a GRB \citep{Sarin2024}.
On the other hand, although the hard X-ray counterpart of FRBs is still not clear \citep{DeLaunay2016,Sakamoto2021}, the recent association between FRB 200428 and the Galactic magnetar SGR~1935+2154 suggests a magnetar as an origin of FRBs \citep{Bochenek2020}.
This observation also indicates that a bright, unknown magnetar flare can be a hard X-ray counterpart of FRBs, which could be identified as a GRB.

\begin{figure}[htbp]
  \begin{center}
    \includegraphics[width=11cm]{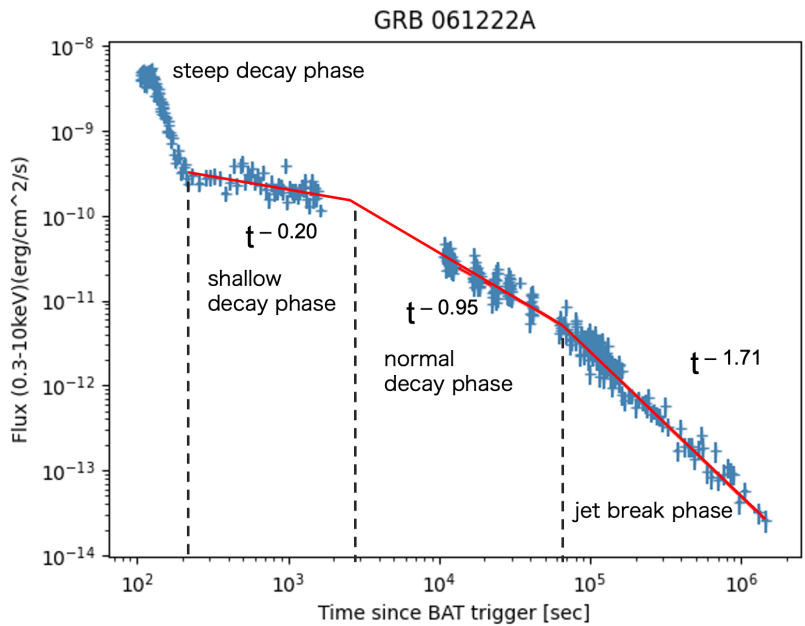}
  \end{center}
  \caption{A typical GRB X-ray afterglow light curve of GRB 061222A.}
  \label{fig:GRB061222A_afterglow}
\end{figure}

One possible scenario of FRBs proposed by \citet{FalckeRezzolla2013} is that a spinning supra-massive neutron star loses centrifugal support and collapses into a black hole. In this case, FRBs would happen several thousand to a million years after the birth of the supra-massive neutron star. On the other hand, \citet{BingZhang2014} suggest that such implosions can happen in supra-massive neutron stars shortly ${\sim} 10^{4}$ s after their births, and the signatures can be visible in X-ray afterglows of some long and short GRBs. X-ray afterglow of GRB shows several different decay phases. Figure \ref{fig:GRB061222A_afterglow} shows one of the typical X-ray afterglow light curves obtained by the \textit{Swift}/X-ray Telescope (XRT) \citep{Burrows2005}. The light curve consists of 4 components: (I) an initial steep decay phase, (II) a shallow decay phase, (III) a normal decay phase, and (IV) a jet break phase. At first, the X-rays decay rapidly in the first few hundred seconds, which is explained as the tail of a prompt emission. After that, the X-ray luminosity attenuates gently for $10^{3} {\sim} 10^{4}$ s. This shallow decay phase requires continuous energy injection into the blast wave \citep{BingZhang2006}, which would be consistent with a spinning-down neutron star engine. At the last part of the X-ray light curves, a normal decay phase with a typical decay index ${\alpha}$ ${\simeq}$ ${-1}$ could be observed. In some GRBs, a further steepening (decay index ${\alpha}$ ${\simeq}$ ${-2}$) is detected after the normal decay phase, which is interpreted as a jet break feature \citep{Rhoads1999}. 

\begin{figure*}[htbp]
  \begin{center}
    \includegraphics[width=16cm]{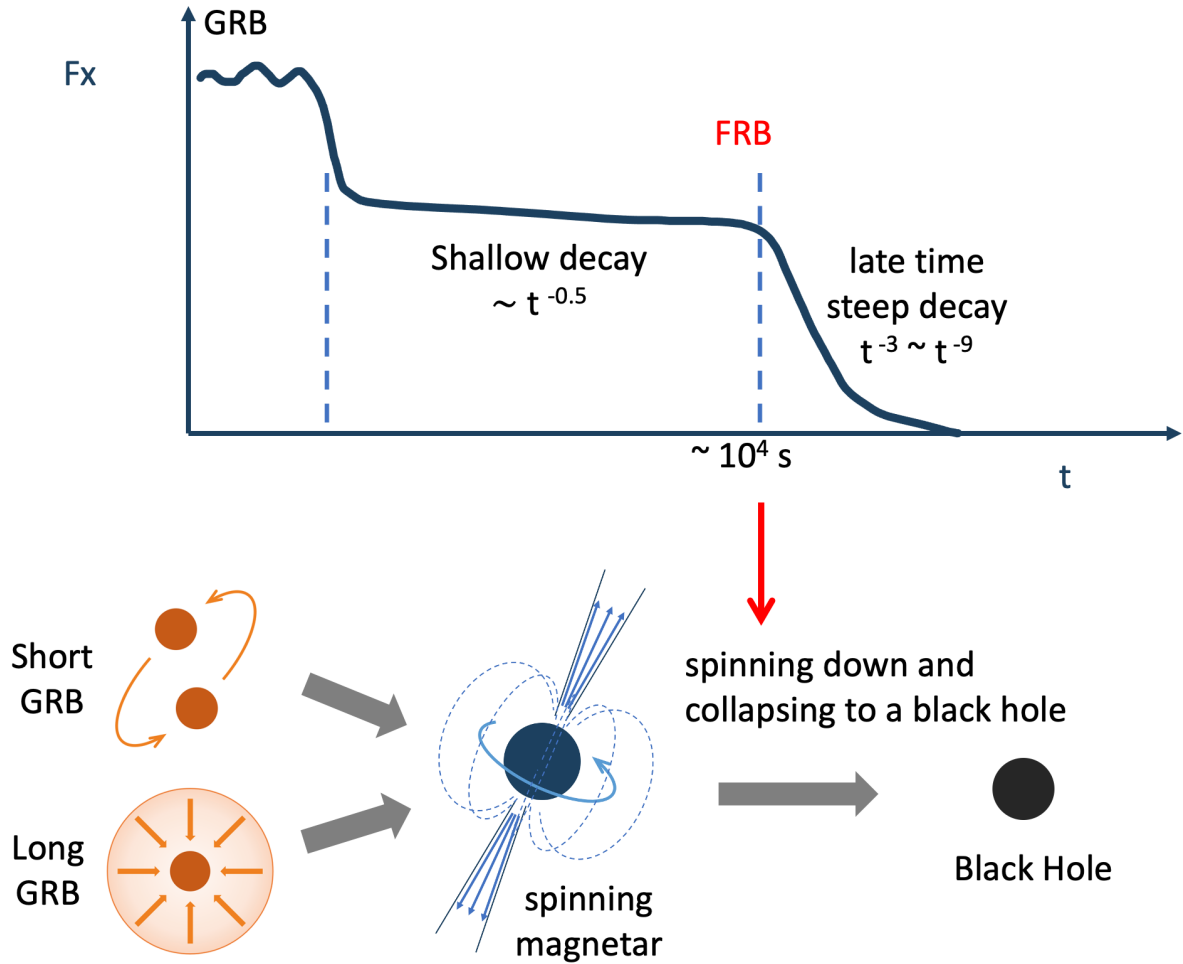}
  \end{center}
  \caption{%
     The model of late time steep decay. When a short or a long GRB occurs, a spinning-down magnetar lasts as the central engine and injects the energy to the jet continuously. Around ${\sim} 10^{4}$ after the onset, the magnetar spins down and collapses into a black hole when it loses centrifugal support. A FRB happens at this epoch when a supra-massive neutron star collapses into a black hole.
}%
  \label{fig:latetime_steep_decay_model}
\end{figure*}

By contrast, in some X-ray afterglow light curves of GRBs, there are X-ray plateaus followed by an extremely steep decay, with a decay index steeper than $-3$, sometimes reaching $-9$ (upper panel of figure \ref{fig:latetime_steep_decay_model}).
Here, we called this phase a late-time steep decay. 
This sudden drop suggests that the emission stops abruptly, and it can happen when a rapidly spinning-down magnetar collapses into a black hole. The epoch could be the emission epoch of the FRB as suggested in \citet{BingZhang2014}.

In this paper, we investigate the possibility of the FRB counterparts as GRBs by using the data of the Neil Gehrels \textit{Swift} Observatory \citep{Gehrels2004}.
Section 2 introduces our search for the event that matches the scenario of \citet{BingZhang2014} and shows how we select and analyze the sample data. We show our results in section 3 and discuss the connection between FRBs and GRBs in section 4. All quoted errors in this work are at the 68\% confidence level.

\section{Observations}
We searched for the X-ray afterglow light curves, which have a late-time steep decay from the {\it Swift}/XRT GRB catalog \citep{Evans2009}. We extracted the time when the decay index suddenly changed by fitting these light curves. Then, we compared the break time with the onset of an FRB to find if there is any associated event between an FRB and a GRB.

\begin{figure*}[htbp]
  \begin{center}
    \includegraphics[width=14cm]{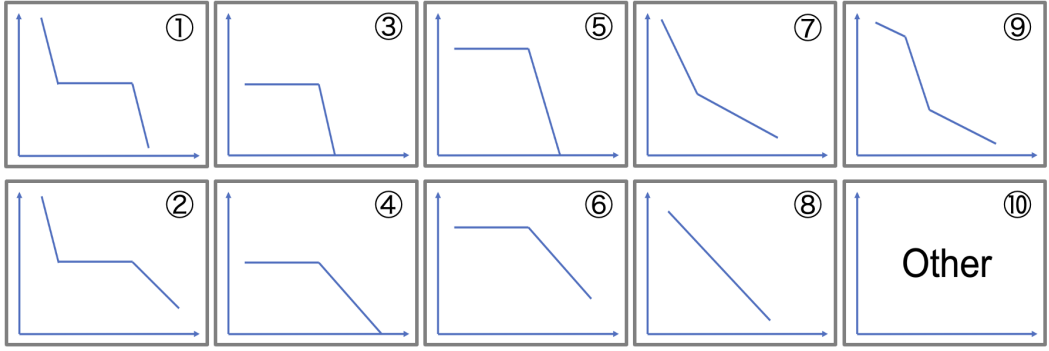}
  \end{center}
  \caption{GRB X-ray afterglow classfication. Type 1 is the light curve with steep ${\rightarrow}$ shallow ${\rightarrow}$ steep (${\alpha} < {-3}$). Type 2 is the type that has a so-called normal decay phase (${\alpha} {\sim} {-1}$) after the shallow decay phase. Type 3 has a steep decay (${\alpha} < {-3}$) following the plateau. By contrast, the decay of type 4 after the plateau is gentler (${\alpha} > {-3}$). Type 5 is similar to type 3, and type 6 is similar to type 4, but the X-ray flux of the plateau is higher than $ 5 \times 10^{-9}$ [erg cm$^{-2}$ s$^{-1}$]. Type 7 first has a steep decay, followed by a gentle slope. Type 8 is the light curve decay in a single power law. Type 9 is shallow ${\rightarrow}$ steep ${\rightarrow}$ shallow. The light curves with too few data points or cannot be classified belong to type 10. Our light curve targets are type 1, type 3, and type 5. Of the selected 88 samples, 37 were type 1, 27 were type 3, and 24 were type 5.
}
  \label{fig:X-ray_afterglow_classfication}
\end{figure*}

\begin{figure*}[htbp]
  \begin{center}
    \includegraphics[width=17cm]{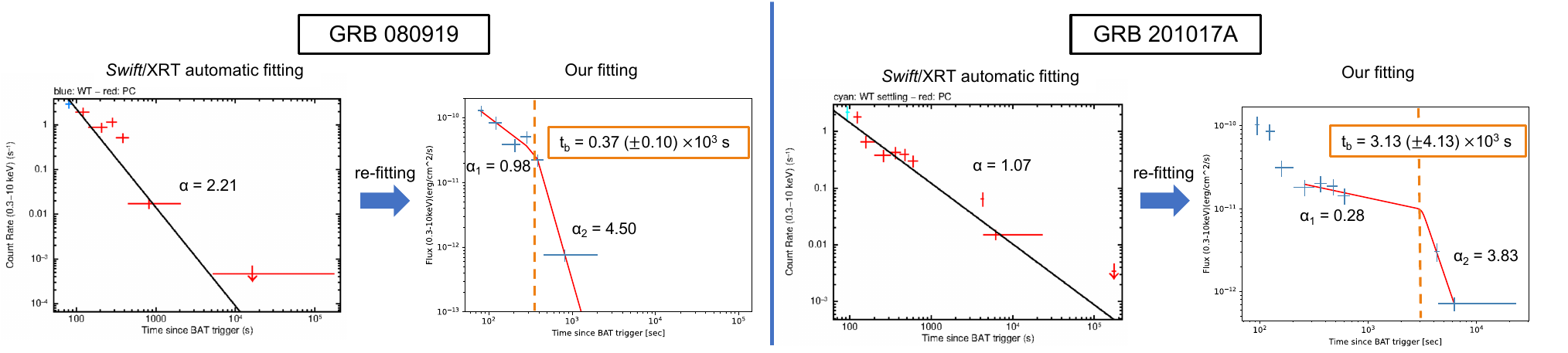}
  \end{center}
  \caption{X-ray afterglow light curves of GRB~080919 (left) and GRB~201017A (riight). For GRB~080919, the decay index ${{\alpha}_2}$ is ${4.50}$ based on our double power-law fitting, and the break time ${t_b}$ is ${0.37 \left(\pm 0.10\right) \times 10^3}$ s. For GRB~201017A, the decay index ${{\alpha}_2}$ is ${3.83}$ based on our double power-law fitting, and the break time ${t_b}$ is ${3.13 \left(\pm 4.13\right) \times 10^3}$ s.}
  \label{fig:04_GRB_fitting_sample}
\end{figure*}

About 1500 X-ray afterglow light curves exist in the {\it Swift}/XRT GRB catalog between 2004 and 2022. We use the following procedure to select our sample to find light curves with a steep decay (temporal decay index is steeper than ${-}$3) after the shallow decay phase. First, we classified all the light curves into ten types by shape, the overall X-ray flux, and the decay index from the automatic light curve fitting parameters in the {\it Swift}/XRT GRB catalog (figure \ref{fig:X-ray_afterglow_classfication}). Then, we picked up the light curves, which have a late-time steep decay. Our targets of the GRBs with a late-time steepening correspond to type 1, type 3, and type 5 in our classification. After the classification, we picked up 86 light curves.
We excluded the same time intervals of X-ray flares identified on the {\it Swift}/XRT GRB catalog, and fitted these light curves to the following double power-law functions \citep{Liang2007},
\begin{equation}\label{double_powerlaw}
F = {F_0} {\left[ {\left(\frac{t}{t_b}\right)}^{{\omega}\,{\alpha}_1} + {\left(\frac{t}{t_b}\right)}^{{\omega\,}{\alpha}_2} \right]}^{-{1/\omega}}, 
\end{equation}
where ${t_b}$ is a break time, ${\omega}$ describes the sharpness of the break, $\alpha_{1}$ is the decay index in the shallow decay phase, and $\alpha_{2}$ is the decay index after ${t_b}$. Here we fixed ${\omega} =$ 10. 
The reason for re-fitting the light curve is to obtain a robust result for our purpose.
Accepting the fit of a double power-law function over a simple power-law, we request the F-test probability of less than 0.15. For instance, figure \ref{fig:04_GRB_fitting_sample} compares the fitting results of two least significant GRBs based on the {\it Swift}/XRT GRB catalog and ours.
As can be seen, for GRB~080919 (figure \ref{fig:04_GRB_fitting_sample} left), the light curve shows a clear steepening in the decay index from $-0.98$ to $-4.50$ as in our fitting, whereas the fitting based on the {\it Swift}/XRT GRB catalog gives a simple power-law as the best-fit function and shows a decay index of $-2.21$. The ${{\chi}^2/d.o.f.}$ for a simple power-law fit is 34/4 while the ${{\chi}^2/d.o.f.}$ for the double power-law fit is 5/2. The F-test probability is 0.13, which makes the double power-law function more significant based on our criterion. As another example, our fitting for GRB~201017A (figure \ref{fig:04_GRB_fitting_sample} right) shows a sudden change in the decay index from $-0.28$ to $-3.83$, whereas the automatic fitting of the {\it Swift}/XRT GRB catalog shows a simple decay index of $-1.07$. The F-test probability between the double power-law fit and the simple power-law fit of GRB~201017A is 0.10.
After re-fitting, we removed the samples that did not meet our requirements based on the fitting result. Our requirement is the late time steep decay index should be steeper than ${-3}$ ($\alpha_{2} >$ 3). If $\alpha_{2}$ is steeper than $-3$ within the error, we included it in our sample to maximize the sample. As a result, we selected 51 light curves as our samples in this paper. Our sample includes 42 long GRBs \citep{Woosley1993} and nine short GRBs \citep{Gehrels2005}.

\section{Results}

\begin{figure*}[htbp]
  \begin{center}
    \includegraphics[width=16cm]{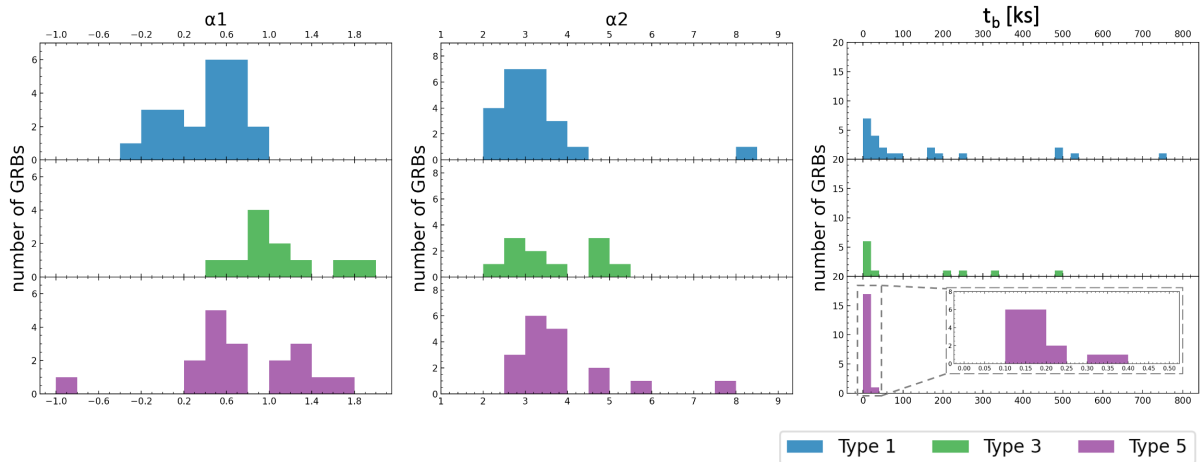}
  \end{center}
  \caption{The histograms of the decay index in the shallow decay phase (${\alpha}_1$), the decay index in the late time steep decay phase(${\alpha}_2$), and the break time(${t_b}$.)}
  \label{fig:a1_a2_tb_hist}
\end{figure*}

Figure \ref{fig:lc_1} shows the XRT light curves of our 51 samples, and table \ref{table1} summarizes the fitting result and the GRB location information from the {\it Swift}/XRT observations. The histograms of the decay index $\alpha_1$ and $\alpha_2$, and the break time from a shallow to a steep decay of each classified type are shown in figure \ref{fig:a1_a2_tb_hist}. 
We compared the break time ${t_b}$ within $\pm$ 1 hour to the onset of reported FRBs (536 FRBs in the CHIME/FRB Catalog 1 \citep{CHIMEFRB_Catlog1} and FRBs detected by other telescopes from the online FRB catalog \citep{FRBCAT}). We find no FRB event matches the time window of ${t_b}$ and the position. The closest coincident event in time is GRB~171209A and FRB~171209, which was detected by the Parks telescope at 20:34:23.5 UT. The FRB happened 24 minutes after ${t_b}$. 
However, since the position difference between GRB 171209A and FRB 171209 is 74$^{\circ}$, those two events are not associated because the position difference is larger than the localization accuracy of the events.

\begin{figure*}[htbp]
  \begin{tabular}{ccc}
  
  \begin{minipage}[c]{0.33\textwidth}
     \centering
     \includegraphics[width=\textwidth, height=4.3cm]{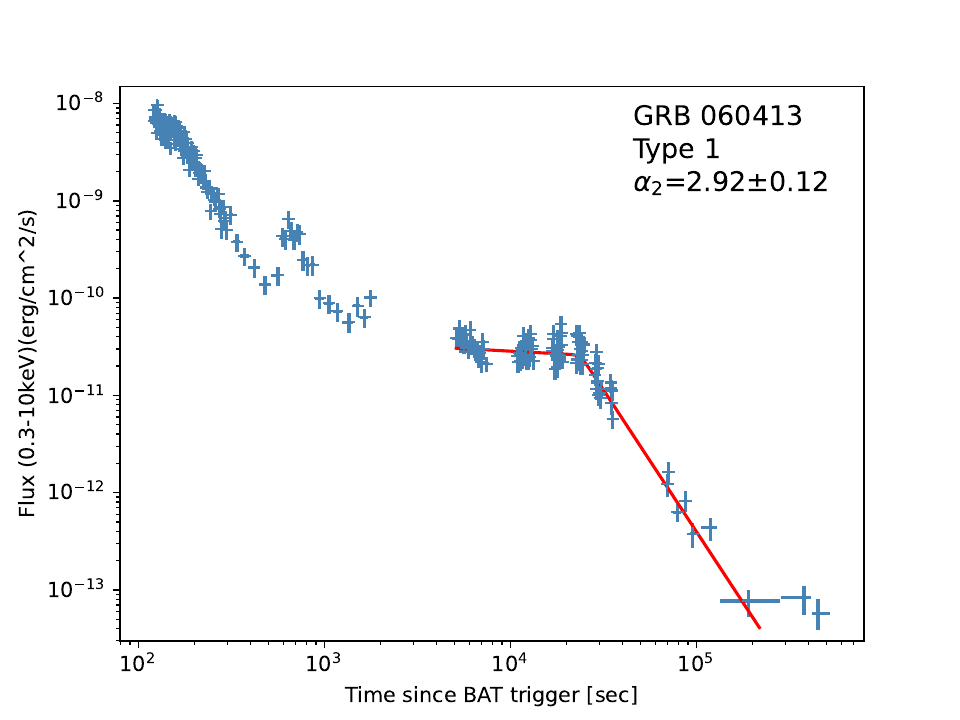}
  \end{minipage}
  \hfill
  \begin{minipage}[c]{0.33\textwidth}
     \centering      
     \includegraphics[width=\textwidth, height=4.3cm]{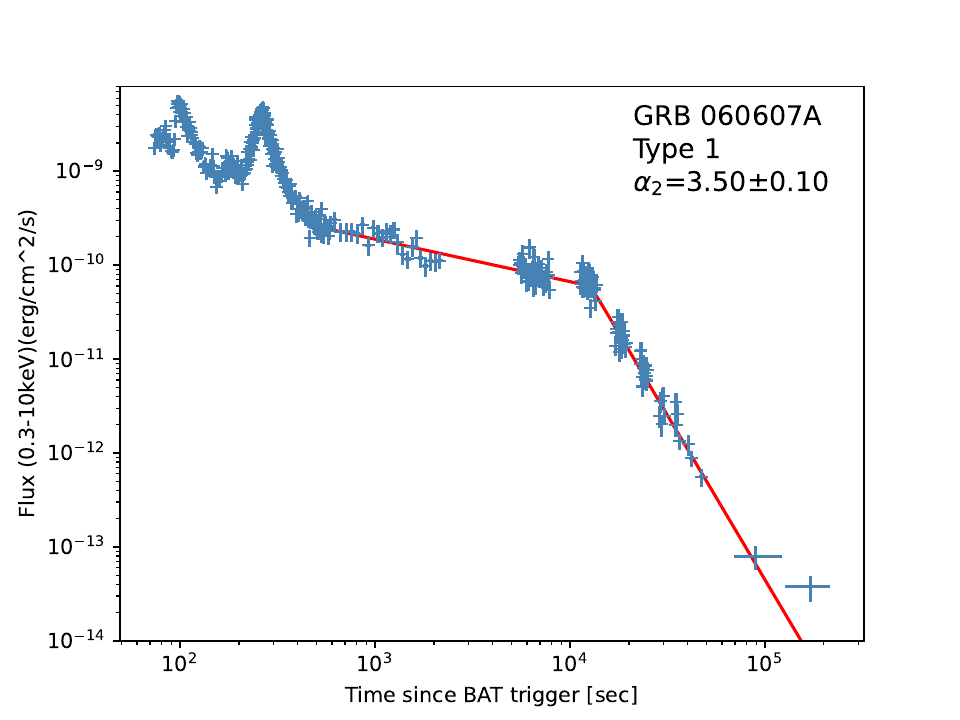}
  \end{minipage}
  \hfill
  \begin{minipage}[c]{0.33\textwidth}
     \centering      
     \includegraphics[width=\textwidth, height=4.3cm]{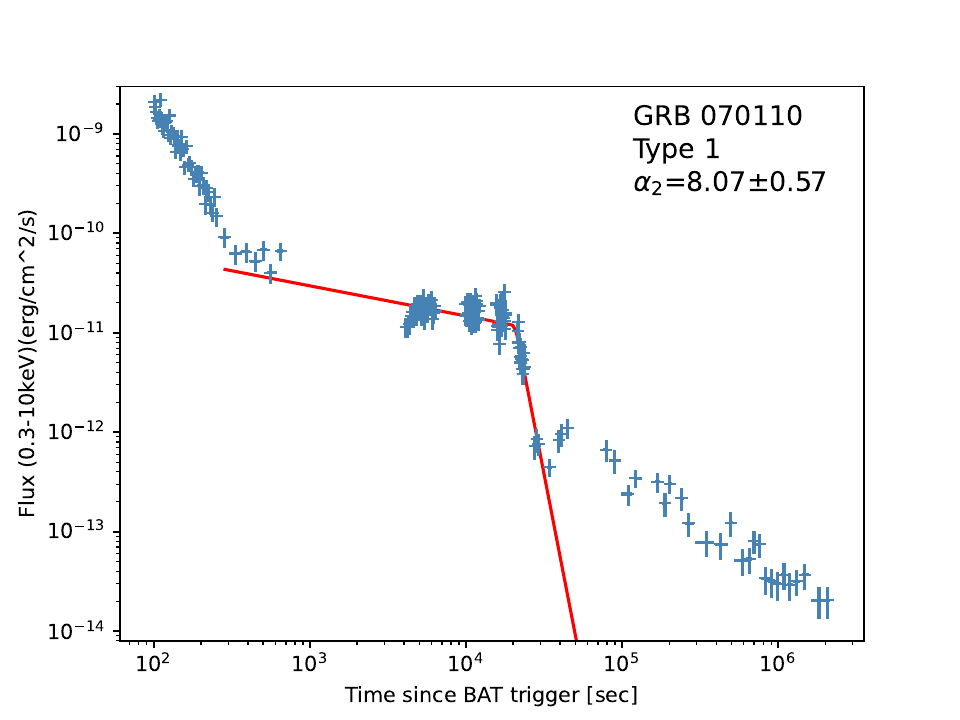}
  \end{minipage}\\

  \begin{minipage}[c]{0.33\textwidth}
     \centering
     \includegraphics[width=\textwidth, height=4.3cm]{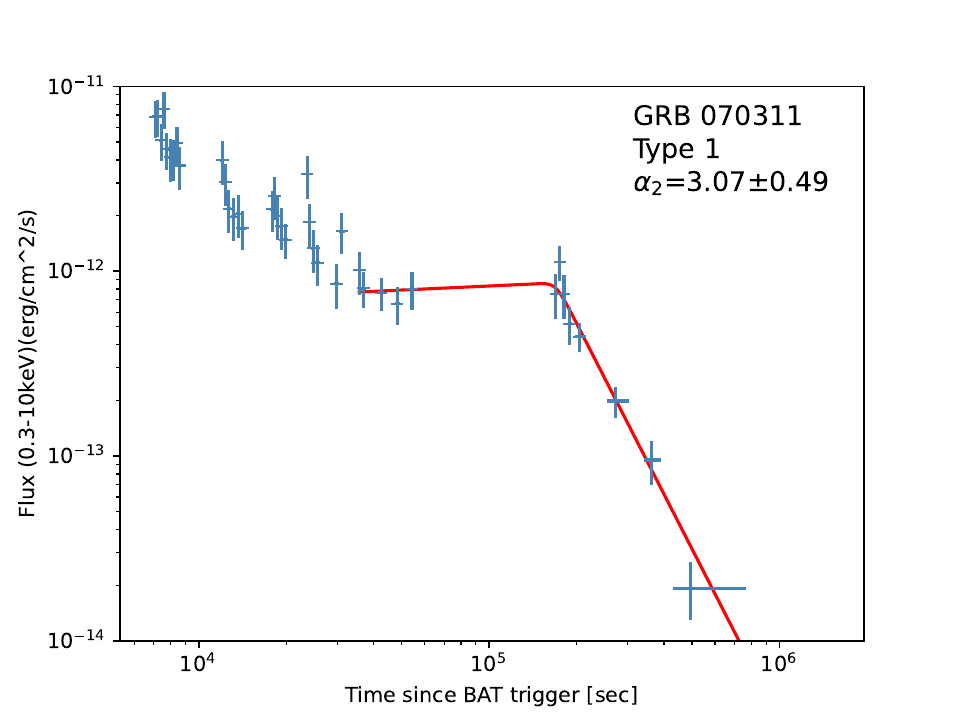}
  \end{minipage}
  \hfill
  \begin{minipage}[c]{0.33\textwidth}
     \centering      
     \includegraphics[width=\textwidth, height=4.3cm]{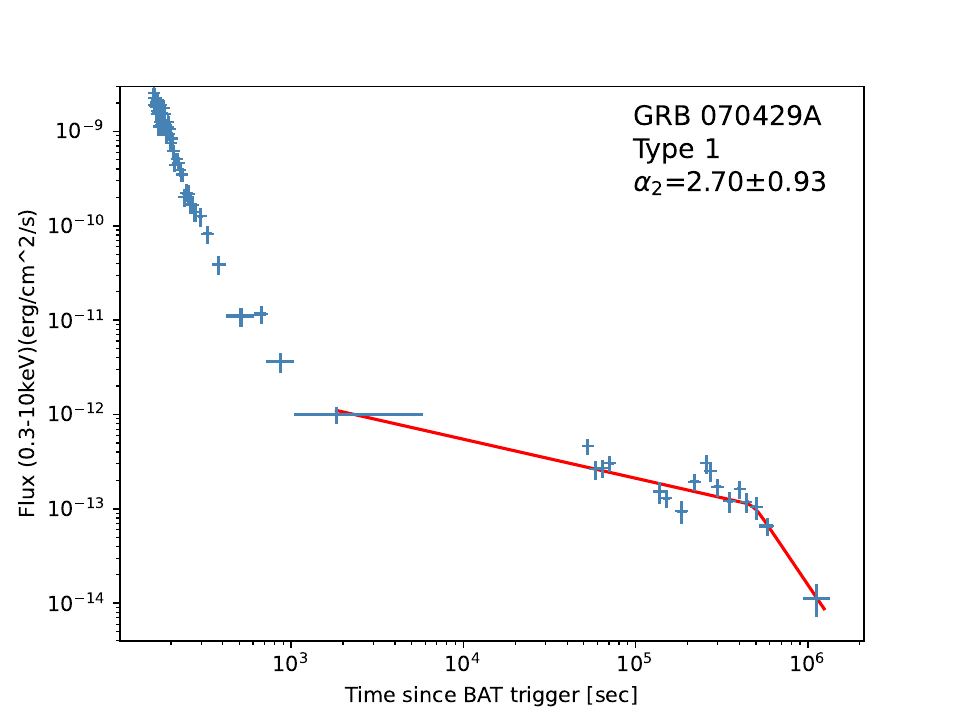}
  \end{minipage}
  \hfill
  \begin{minipage}[c]{0.33\textwidth}
     \centering      
     \includegraphics[width=\textwidth, height=4.3cm]{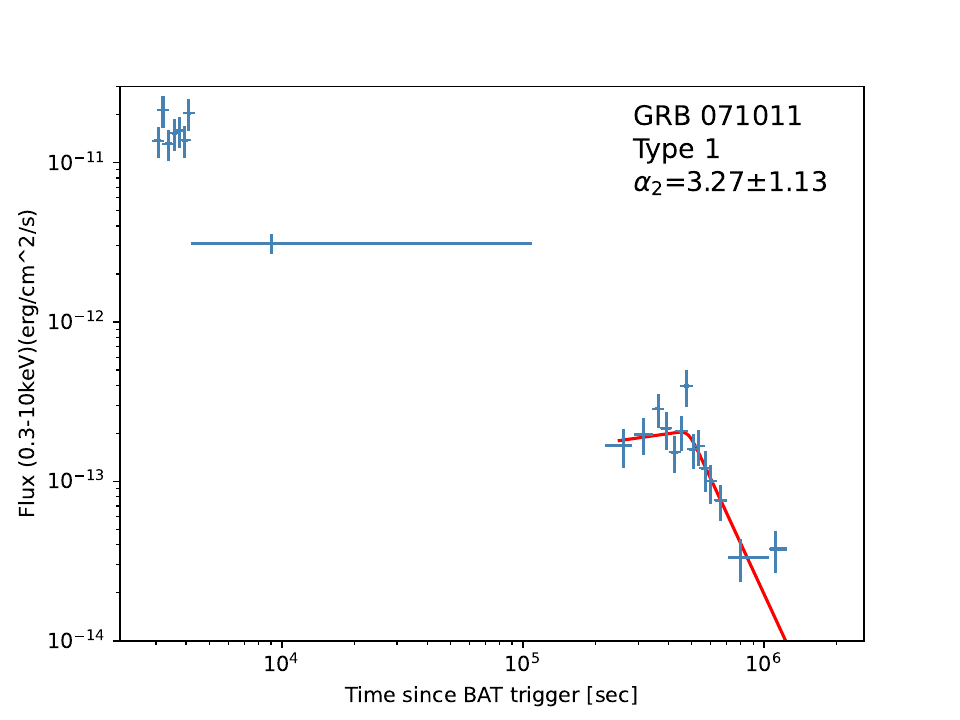}
  \end{minipage}\\

  \begin{minipage}[c]{0.33\textwidth}
     \centering
     \includegraphics[width=\textwidth, height=4.3cm]{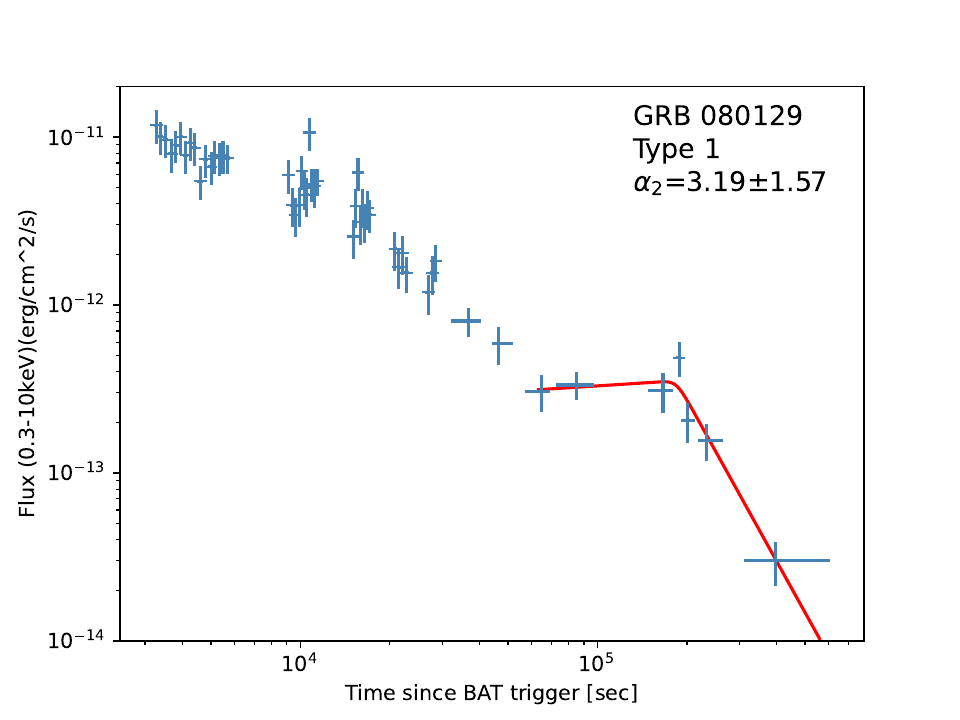}
     \end{minipage}
  \hfill
  \begin{minipage}[c]{0.33\textwidth}
     \centering      
     \includegraphics[width=\textwidth, height=4.3cm]{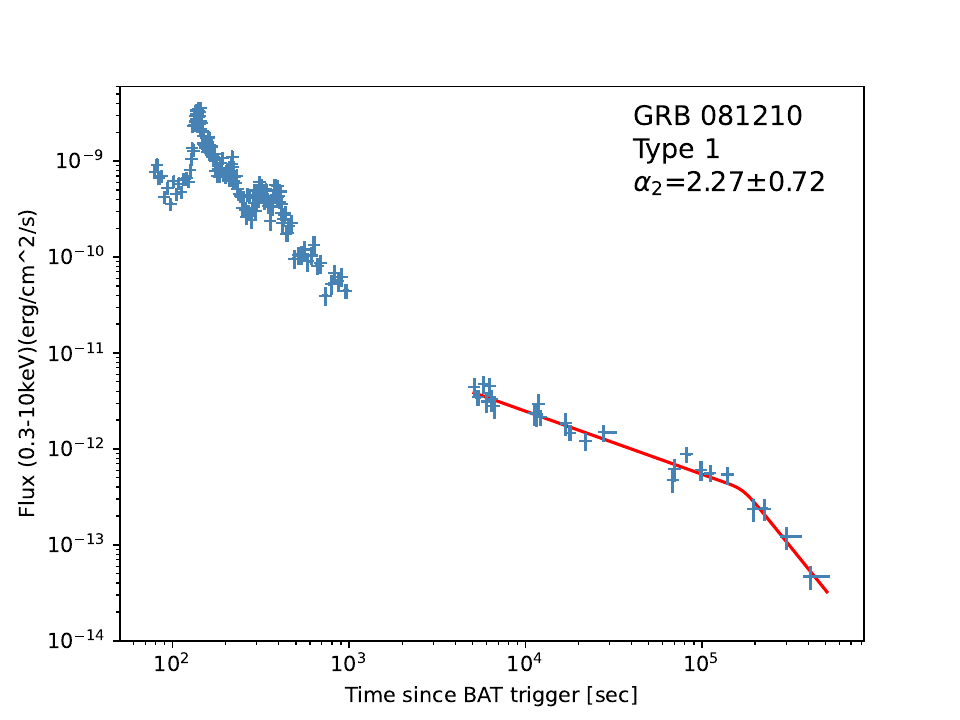}
  \end{minipage}
  \hfill
  \begin{minipage}[c]{0.33\textwidth}
     \centering      
     \includegraphics[width=\textwidth, height=4.3cm]{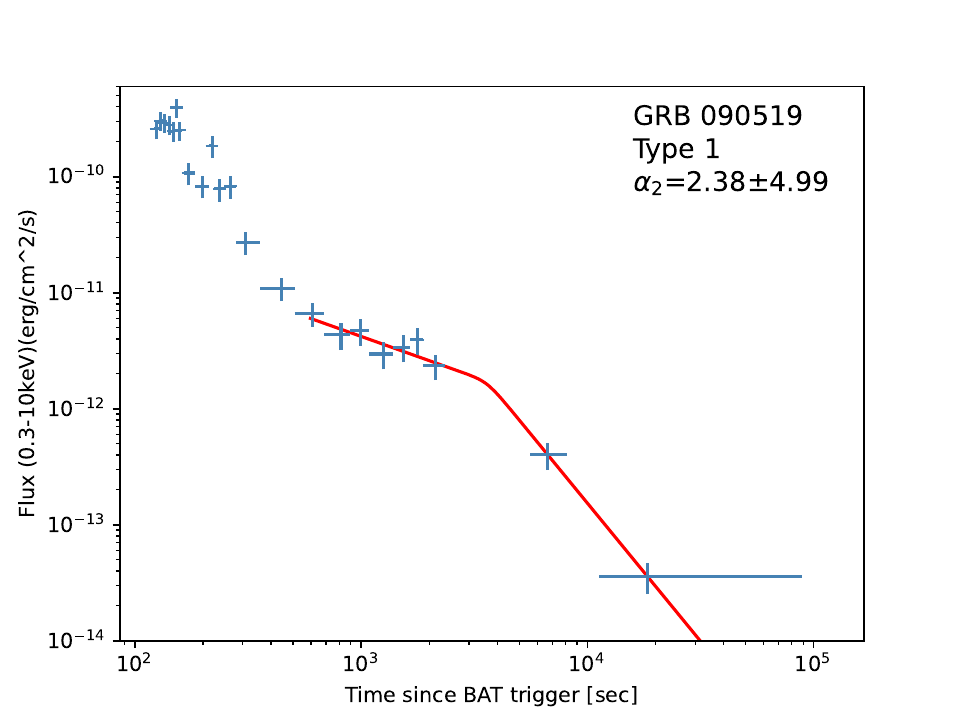}
  \end{minipage}\\

  \begin{minipage}[c]{0.33\textwidth}
      \centering
      \includegraphics[width=\textwidth, height=4.3cm]{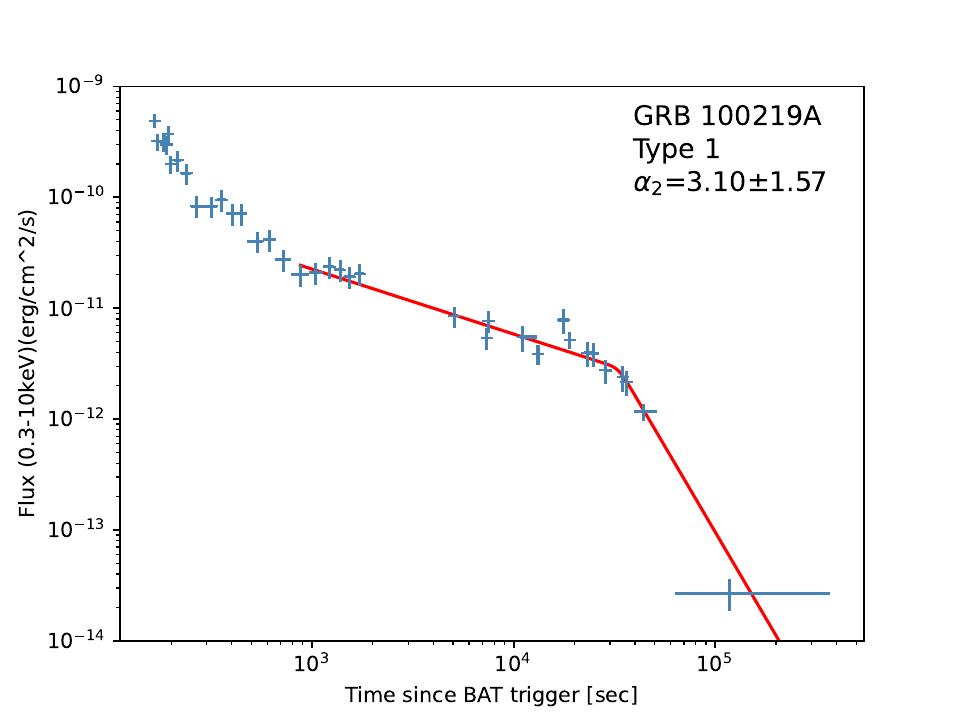}
   \end{minipage}
   \hfill
   \begin{minipage}[c]{0.33\textwidth}
      \centering      
      \includegraphics[width=\textwidth, height=4.3cm]{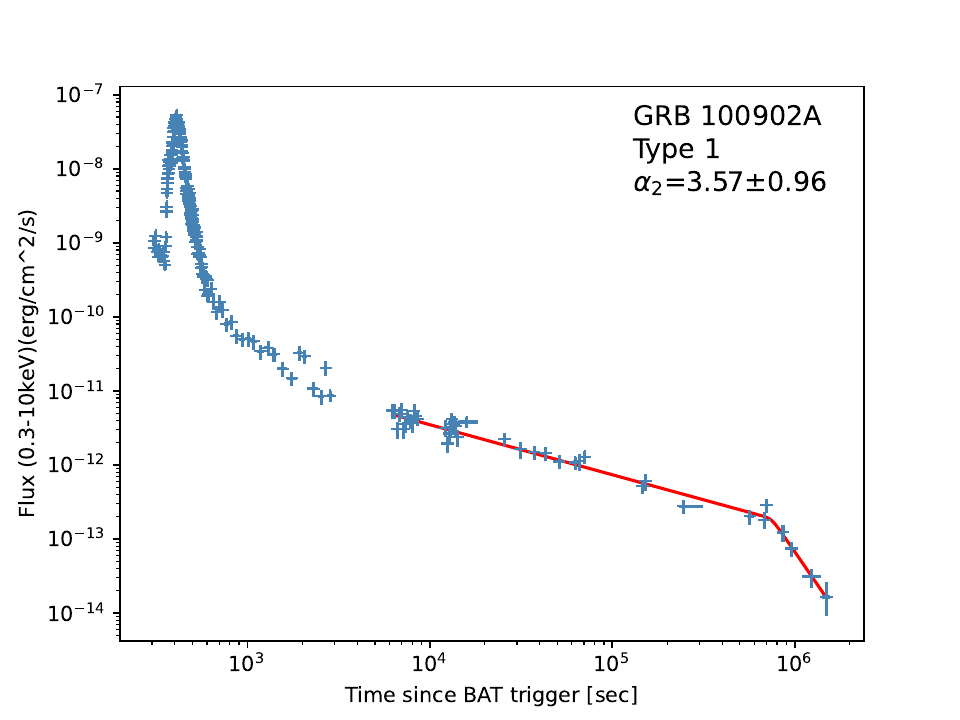}
   \end{minipage}
   \hfill
   \begin{minipage}[c]{0.33\textwidth}
      \centering      
      \includegraphics[width=\textwidth, height=4.3cm]{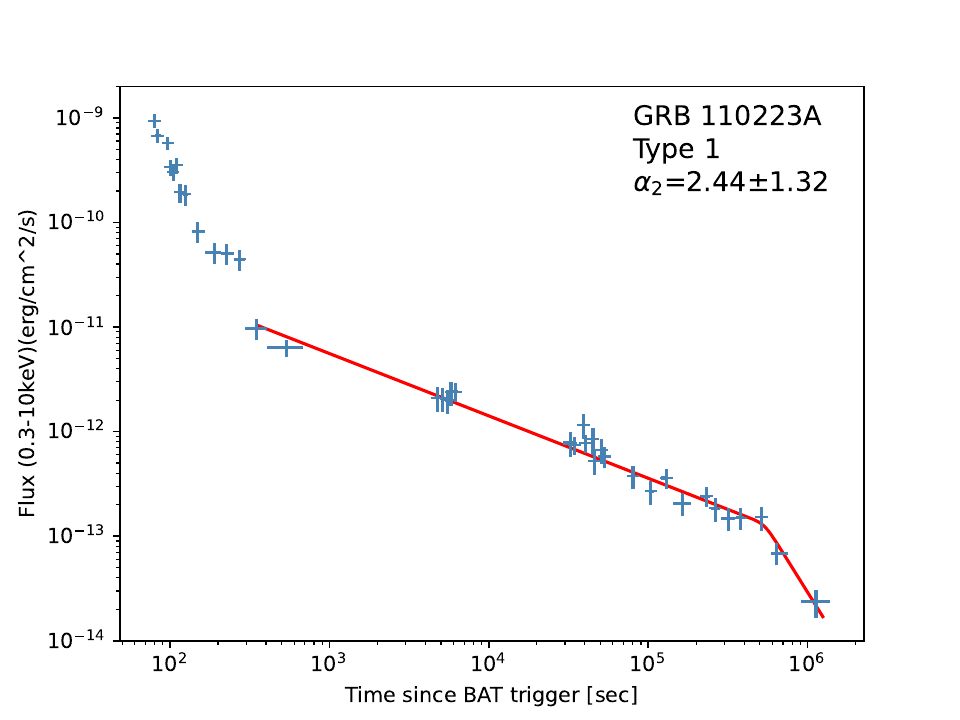}

   \end{minipage}\\

  \end{tabular}
  \caption{X-ray afterglow light curves. The red solid lines best fit a smoothly broken power law from the shallow decay phase to the late time steep decay phase. Each panel shows the classification and decay index in the top right corner.}
  \label{fig:lc_1}
\end{figure*}

\begin{figure*}[htbp]
  \addtocounter{figure}{-1}
  \begin{tabular}{ccc}

   \begin{minipage}[c]{0.33\textwidth}
     \centering
     \includegraphics[width=\textwidth, height=4.3cm]{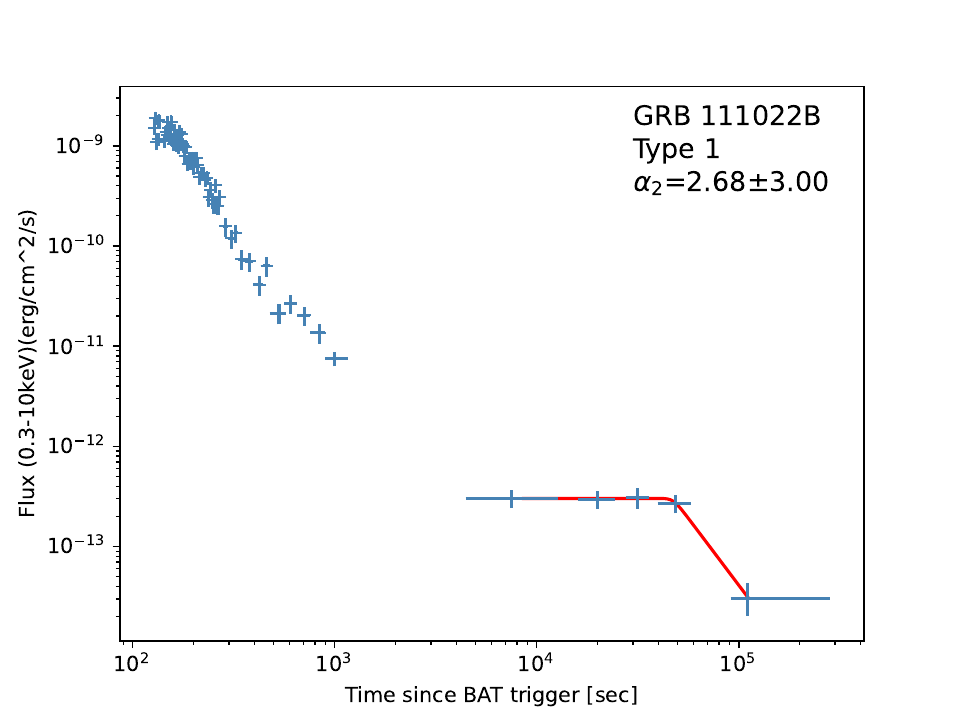}
  \end{minipage}
  \hfill
  \begin{minipage}[c]{0.33\textwidth}
     \centering      
     \includegraphics[width=\textwidth, height=4.3cm]{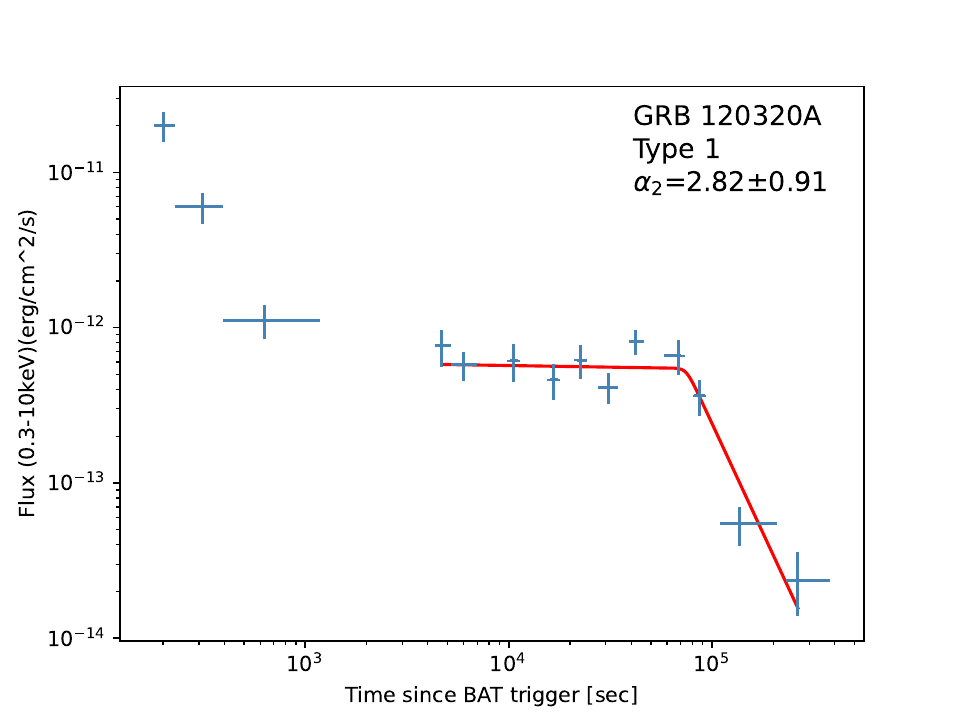}
  \end{minipage}
  \hfill
  \begin{minipage}[c]{0.33\textwidth}
     \centering      
     \includegraphics[width=\textwidth, height=4.3cm]{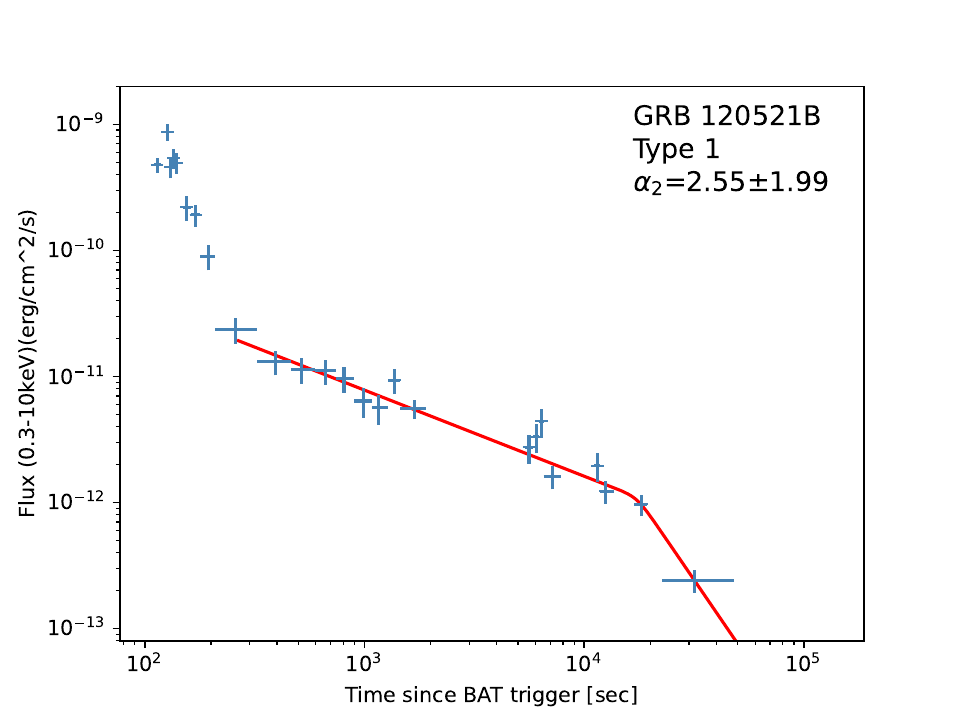}
  \end{minipage}\\

  \begin{minipage}[c]{0.33\textwidth}
     \centering
     \includegraphics[width=\textwidth, height=4.3cm]{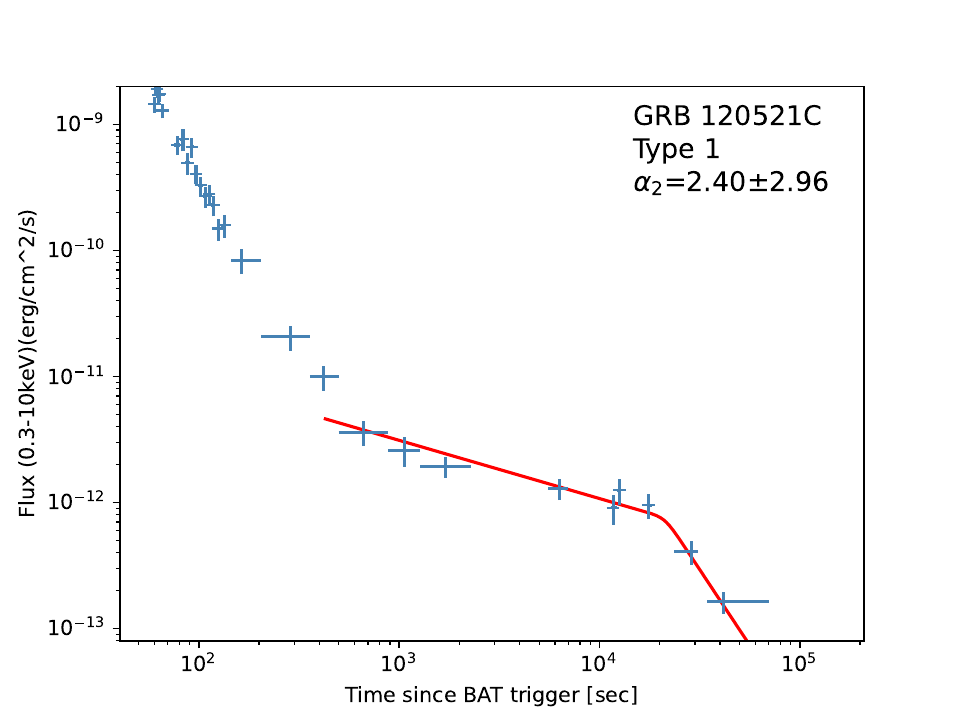}
  \end{minipage}
  \hfill
  \begin{minipage}[c]{0.33\textwidth}
     \centering      
     \includegraphics[width=\textwidth, height=4.3cm]{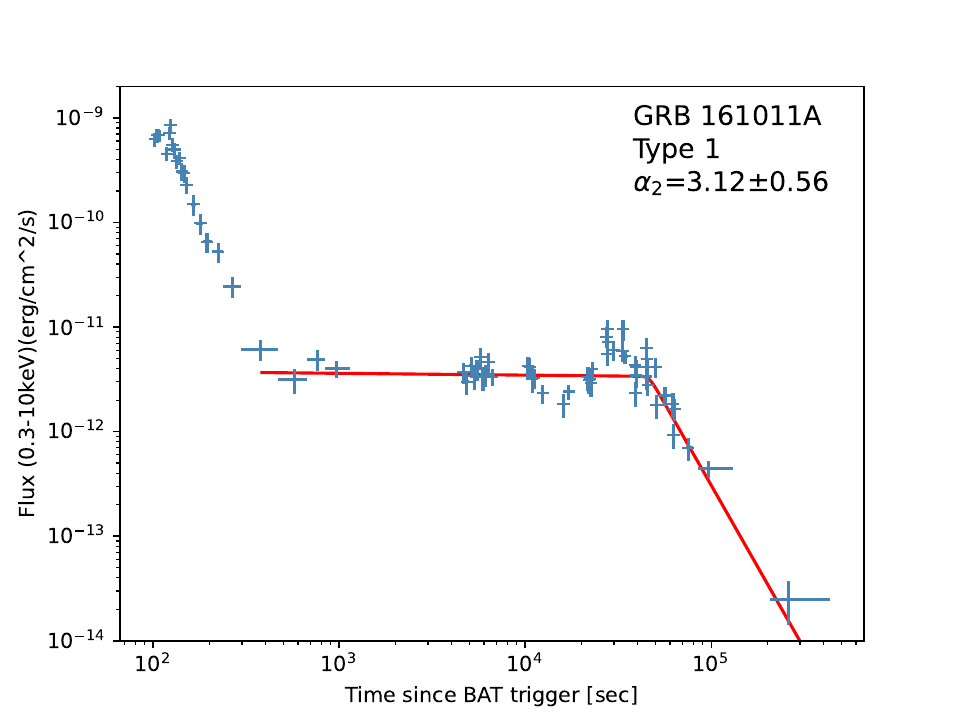}
  \end{minipage}
  \hfill
  \begin{minipage}[c]{0.33\textwidth}
     \centering      
     \includegraphics[width=\textwidth, height=4.3cm]{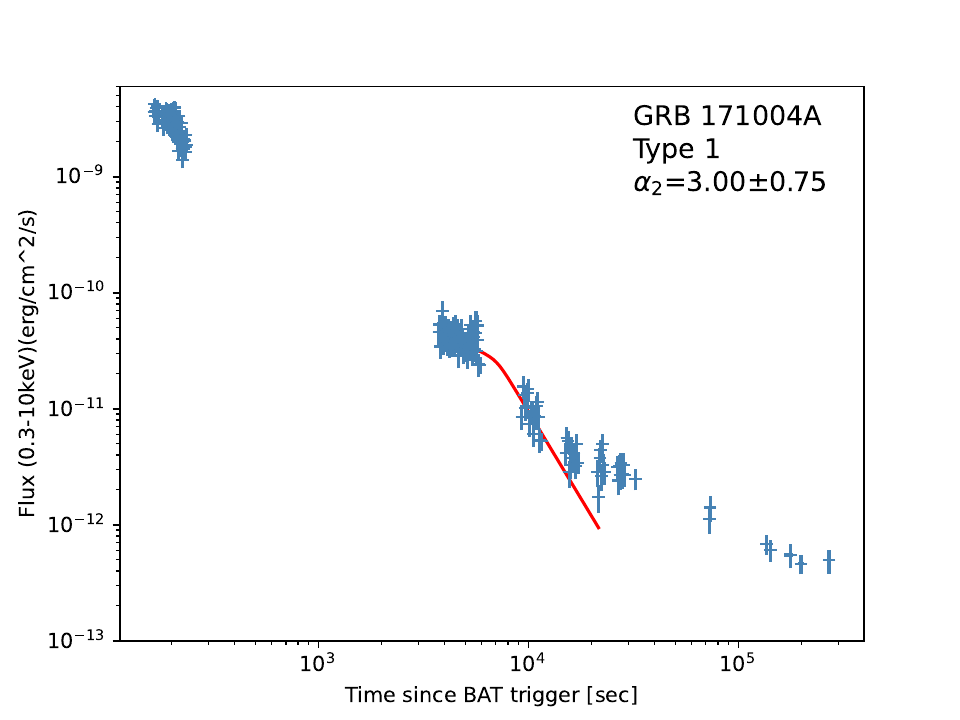}
  \end{minipage}\\

  \begin{minipage}[c]{0.33\textwidth}
     \centering
     \includegraphics[width=\textwidth, height=4.3cm]{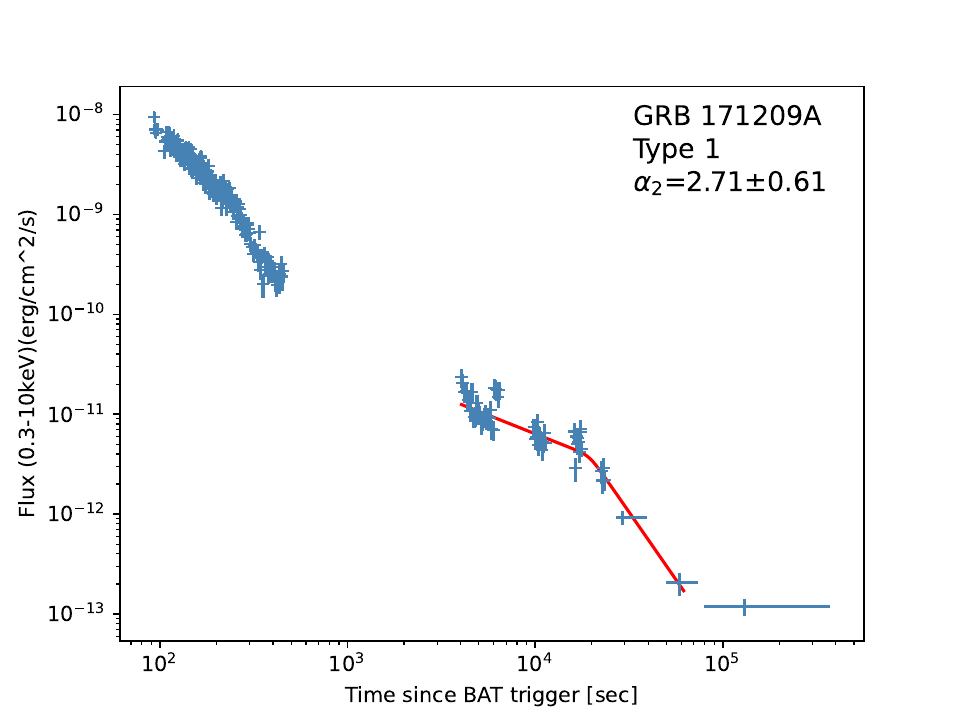}
  \end{minipage}
  \hfill
  \begin{minipage}[c]{0.33\textwidth}
     \centering      
     \includegraphics[width=\textwidth, height=4.3cm]{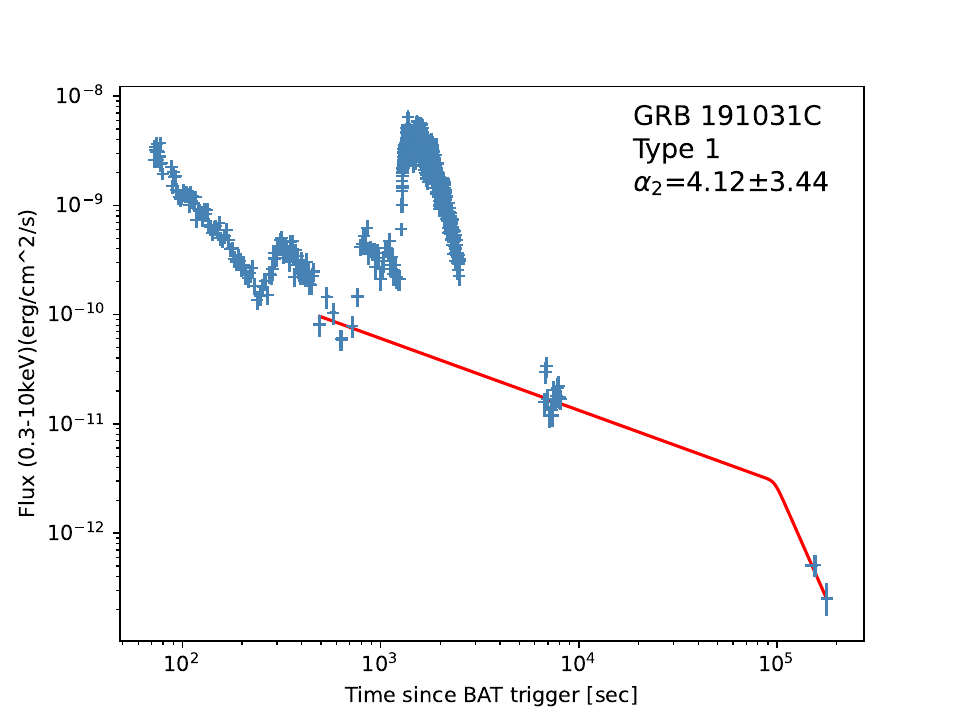}
  \end{minipage}
  \hfill
  \begin{minipage}[c]{0.33\textwidth}
     \centering      
     \includegraphics[width=\textwidth, height=4.3cm]{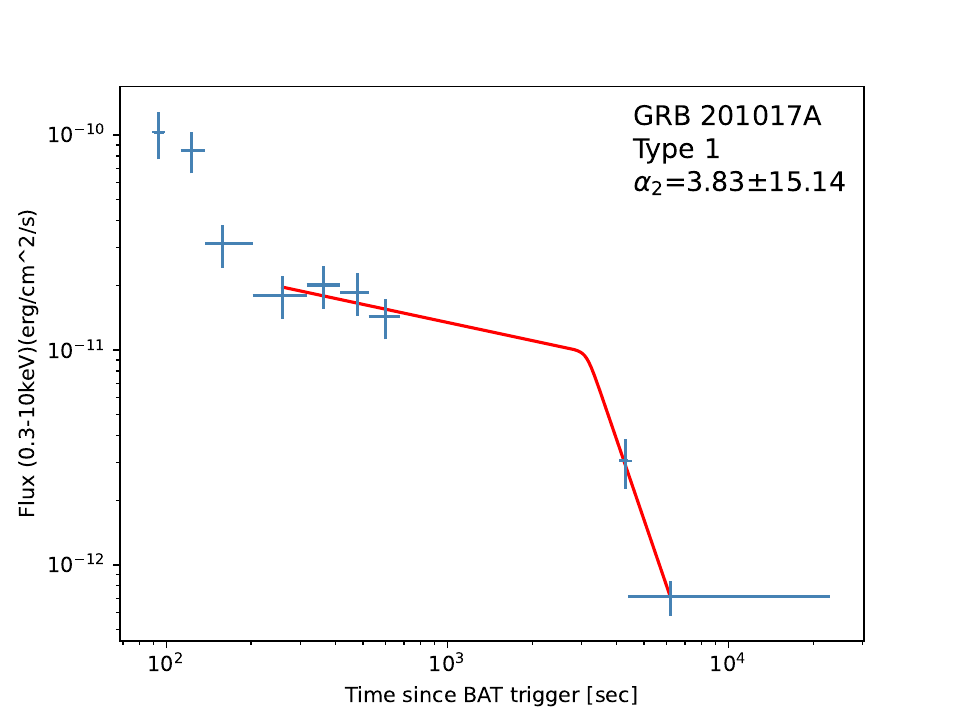}
  \end{minipage}\\

  \begin{minipage}[c]{0.33\textwidth}
     \centering
     \includegraphics[width=\textwidth, height=4.3cm]{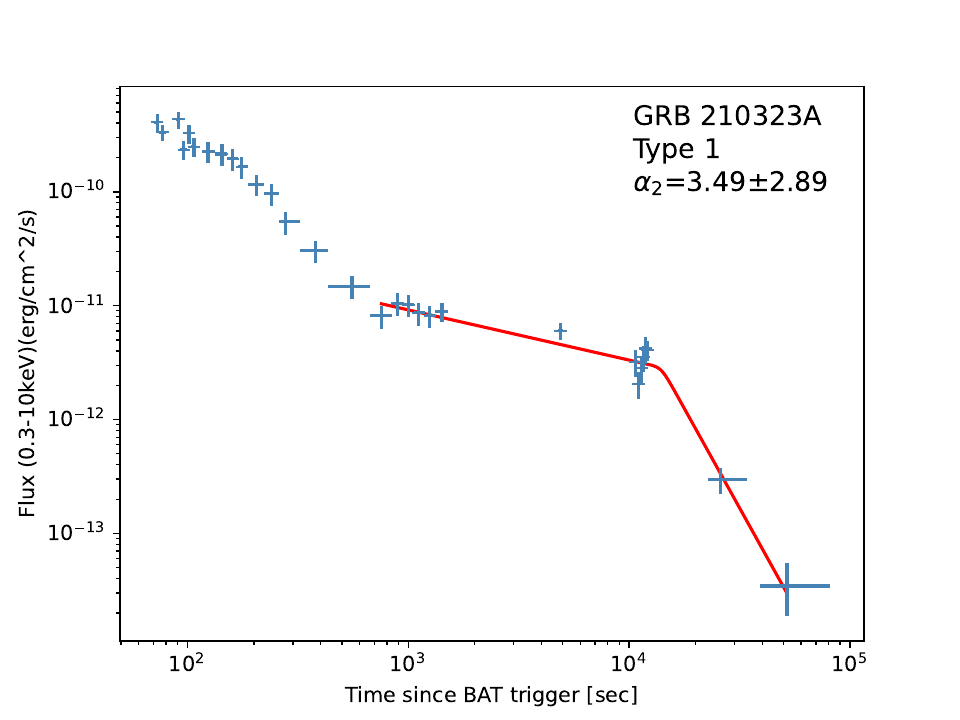}
  \end{minipage}
  \hfill
  \begin{minipage}[c]{0.33\textwidth}
     \centering      
     \includegraphics[width=\textwidth, height=4.3cm]{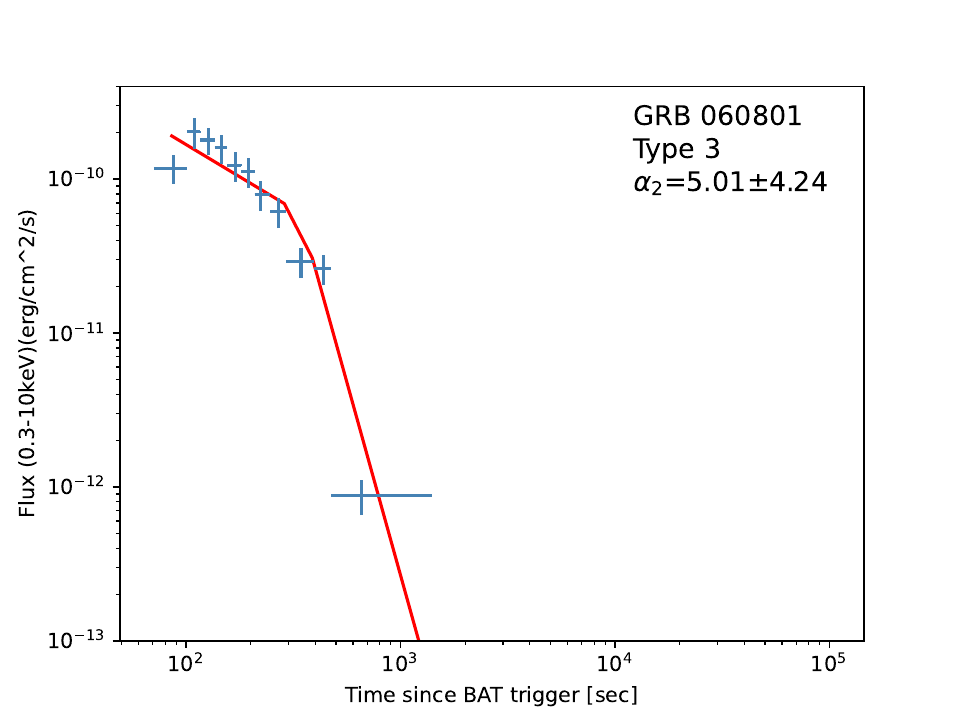}
  \end{minipage}
  \hfill
  \begin{minipage}[c]{0.33\textwidth}
     \centering      
     \includegraphics[width=\textwidth, height=4.3cm]{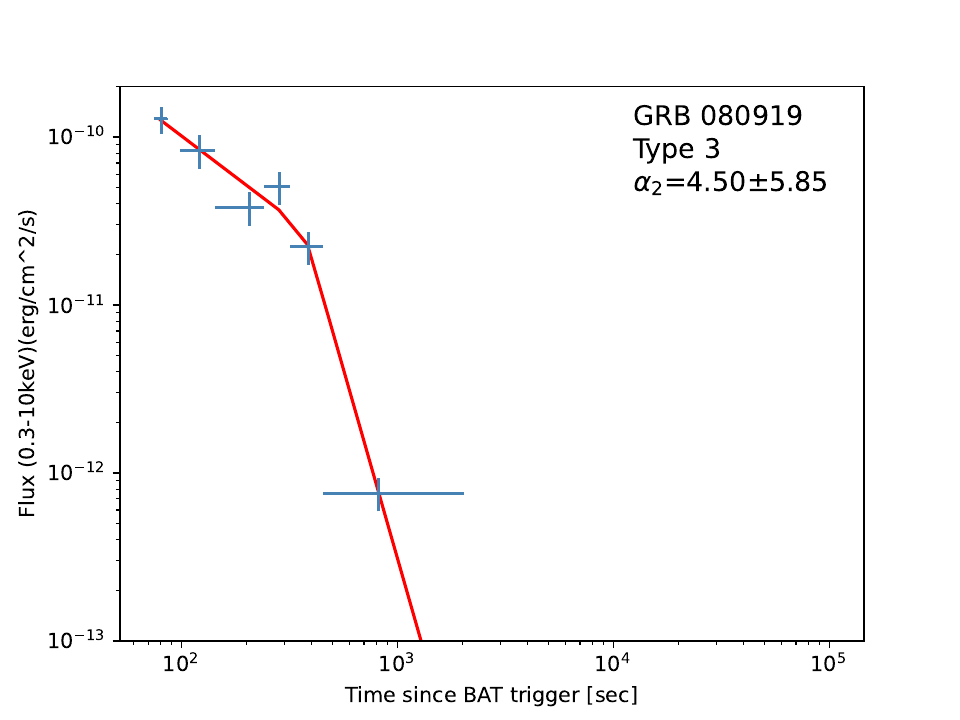}
  \end{minipage}\\

  \begin{minipage}[c]{0.33\textwidth}
      \centering
      \includegraphics[width=\textwidth, height=4.3cm]{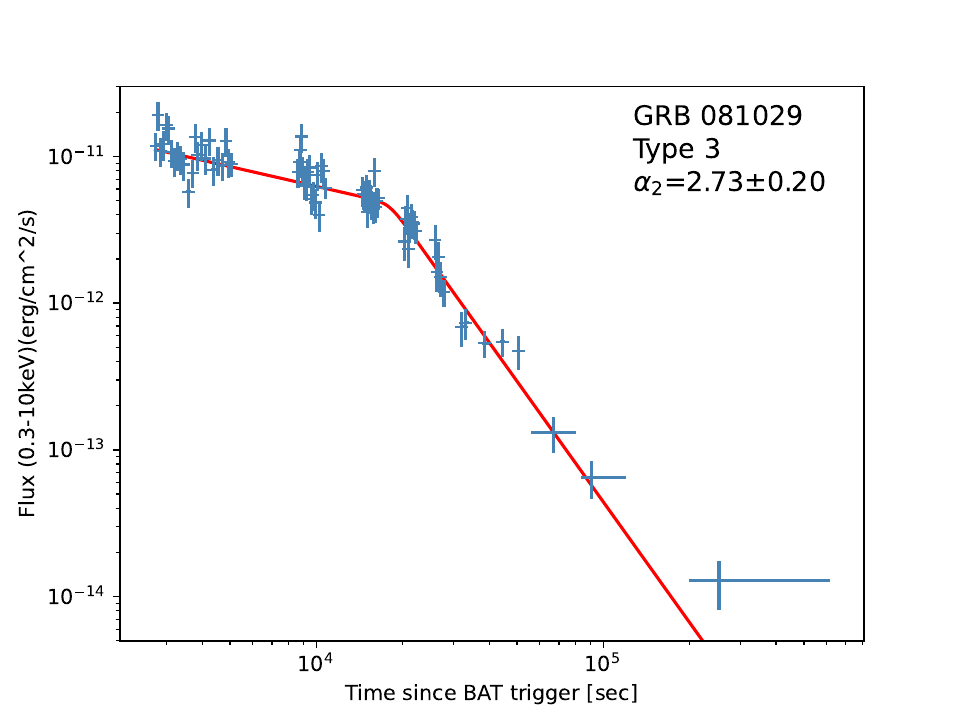}
   \end{minipage}
   \hfill
   \begin{minipage}[c]{0.33\textwidth}
      \centering      
      \includegraphics[width=\textwidth, height=4.3cm]{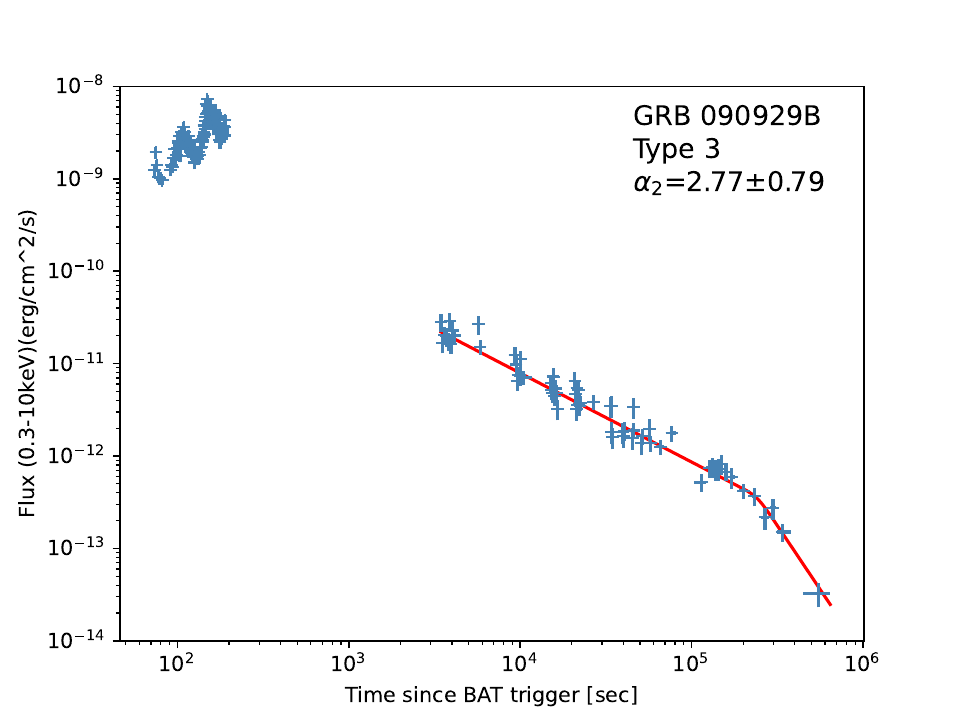}
   \end{minipage}
   \hfill
   \begin{minipage}[c]{0.33\textwidth}
      \centering      
      \includegraphics[width=\textwidth, height=4.3cm]{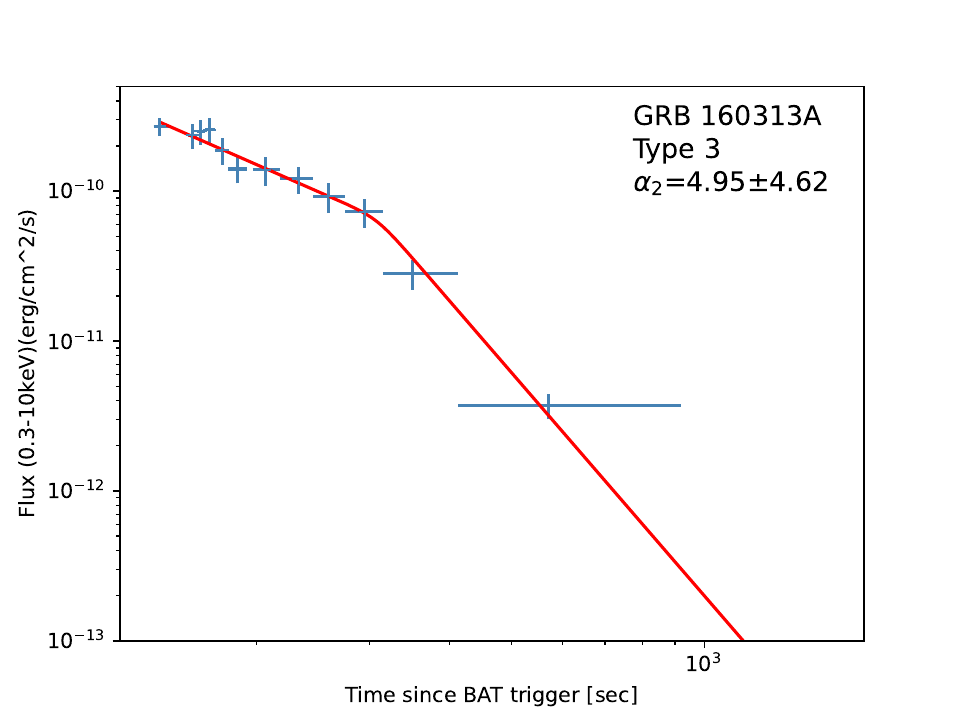}
   \end{minipage}\\

  \end{tabular}
  \caption{Continued: X-ray afterglow light curves.}
  \label{fig:lc_2}
\end{figure*}

\begin{figure*}[htbp]
  \addtocounter{figure}{-1}
  \begin{tabular}{ccc}

  \begin{minipage}[c]{0.33\textwidth}
     \centering
     \includegraphics[width=\textwidth, height=4.3cm]{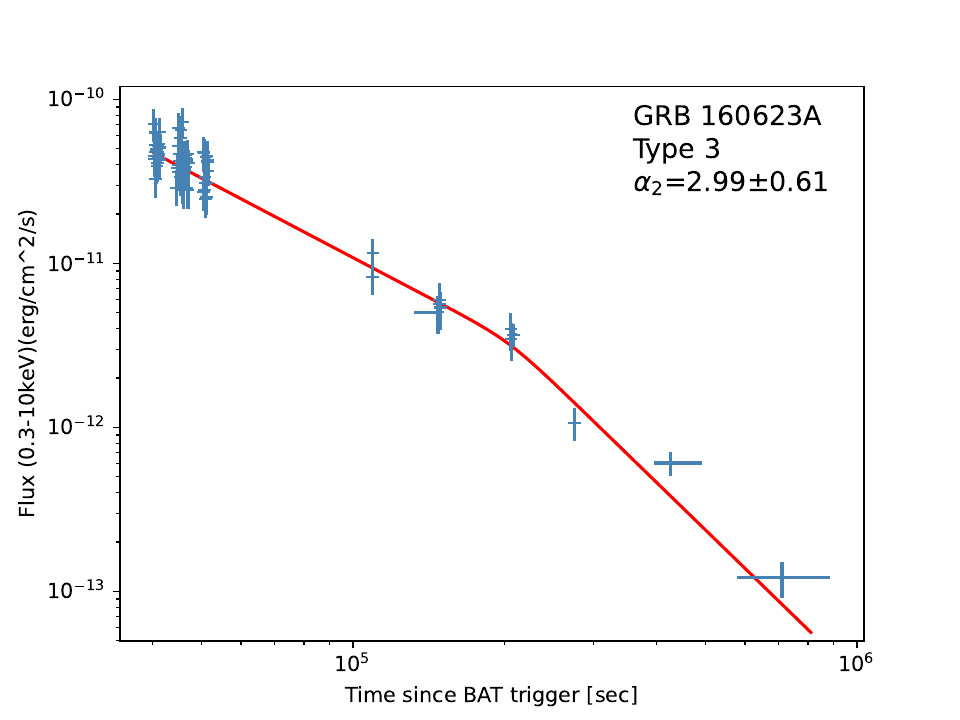}
  \end{minipage}
  \hfill
  \begin{minipage}[c]{0.33\textwidth}
     \centering      
     \includegraphics[width=\textwidth, height=4.3cm]{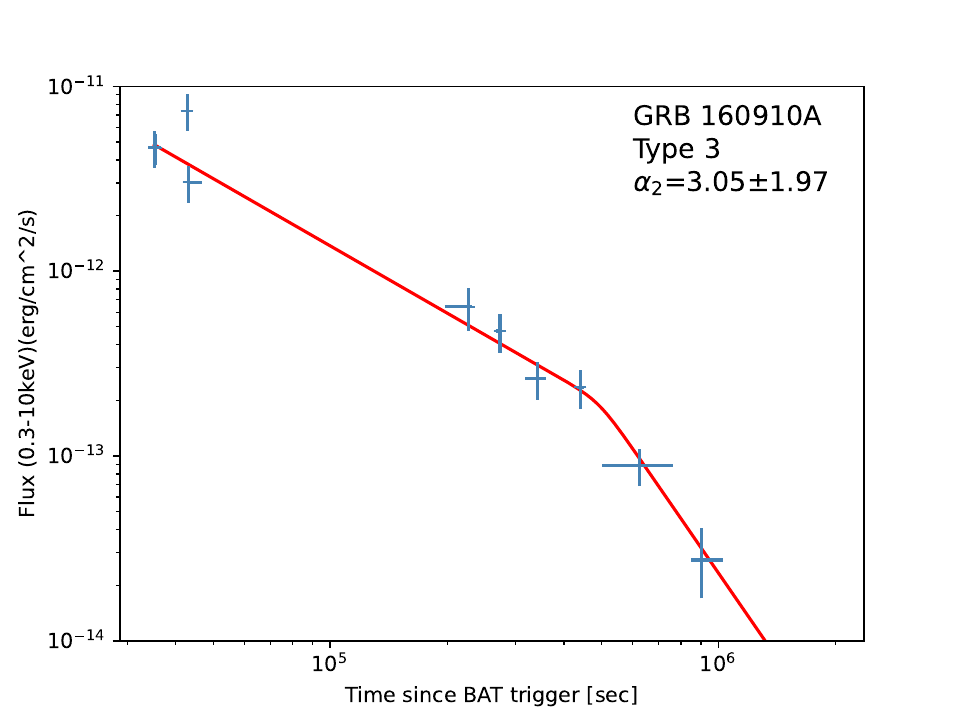}
  \end{minipage}
  \hfill
  \begin{minipage}[c]{0.33\textwidth}
     \centering      
     \includegraphics[width=\textwidth, height=4.3cm]{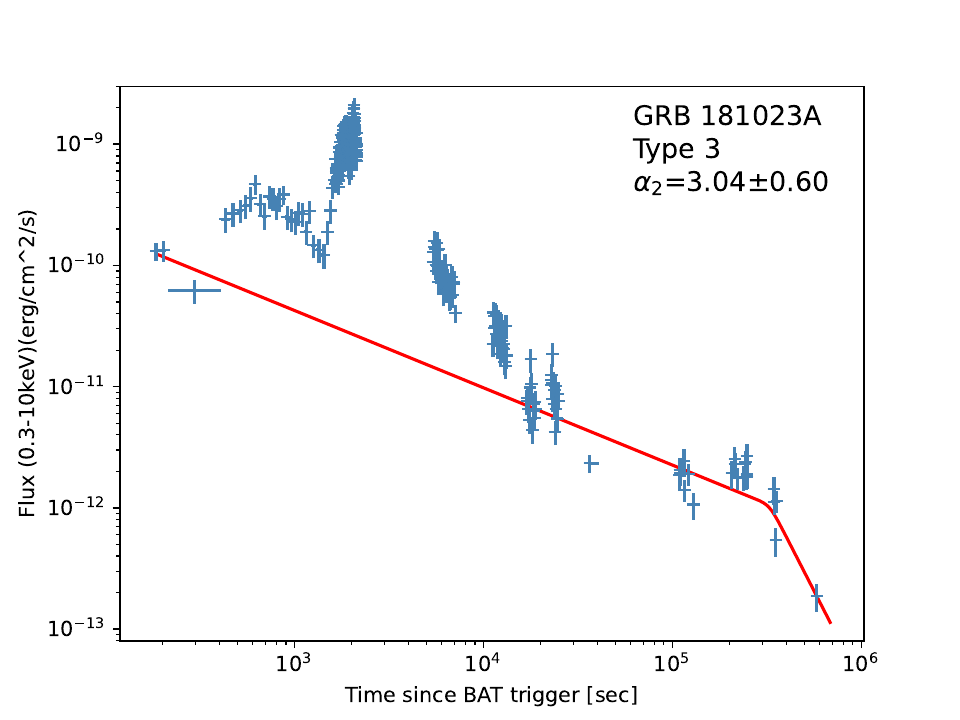}
  \end{minipage}\\
  
  \begin{minipage}[c]{0.33\textwidth}
     \centering
     \includegraphics[width=\textwidth, height=4.3cm]{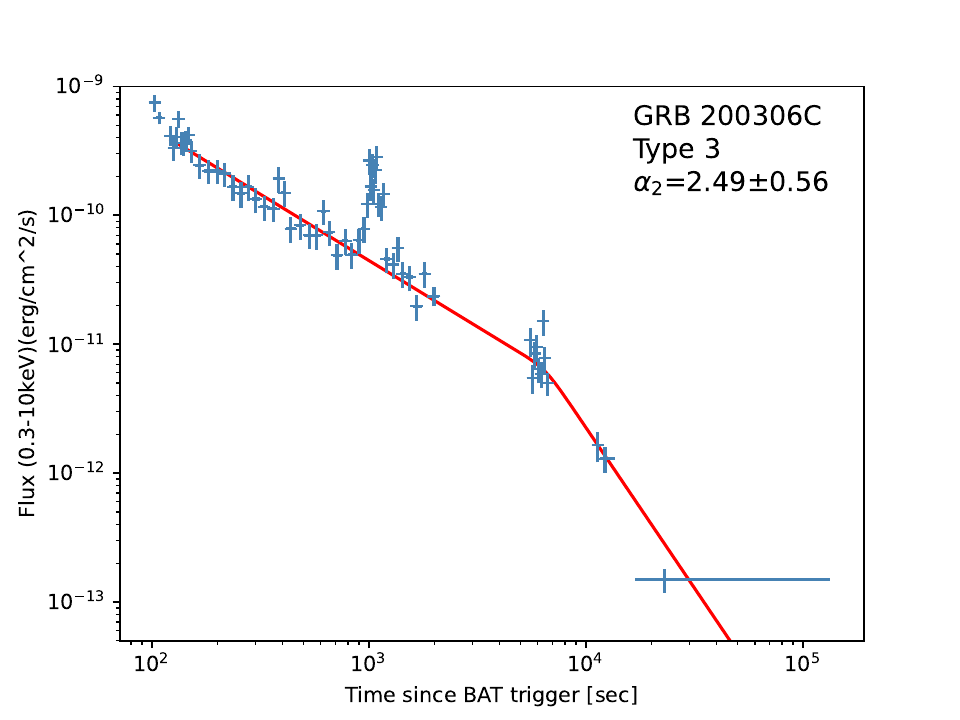}
  \end{minipage}
  \hfill
  \begin{minipage}[c]{0.33\textwidth}
     \centering      
     \includegraphics[width=\textwidth, height=4.3cm]{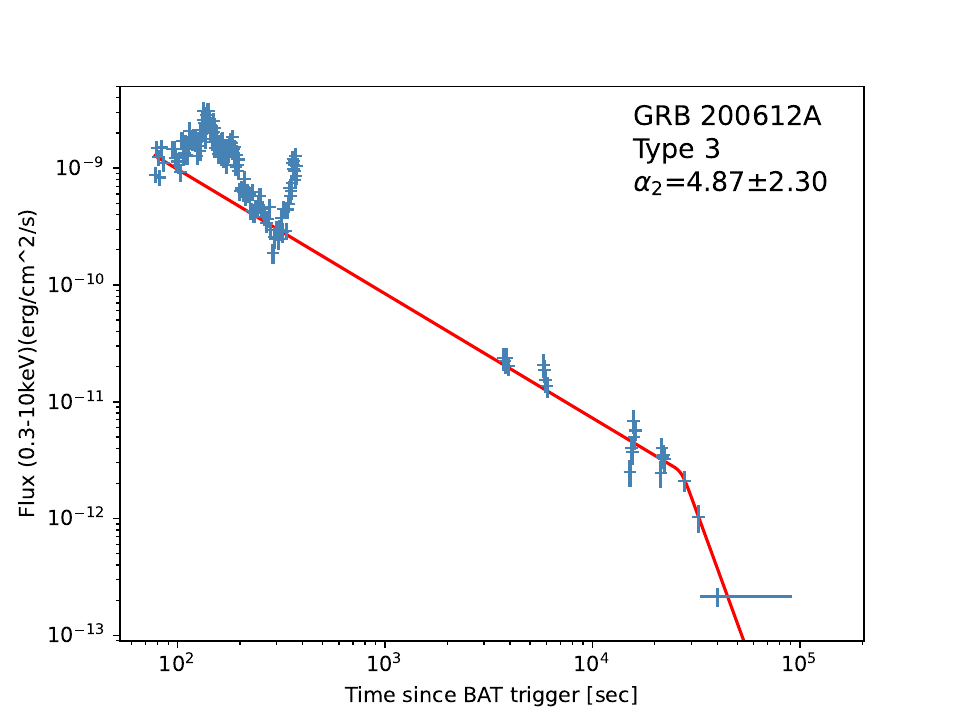}
  \end{minipage}
  \hfill
  \begin{minipage}[c]{0.33\textwidth}
     \centering      
     \includegraphics[width=\textwidth, height=4.3cm]{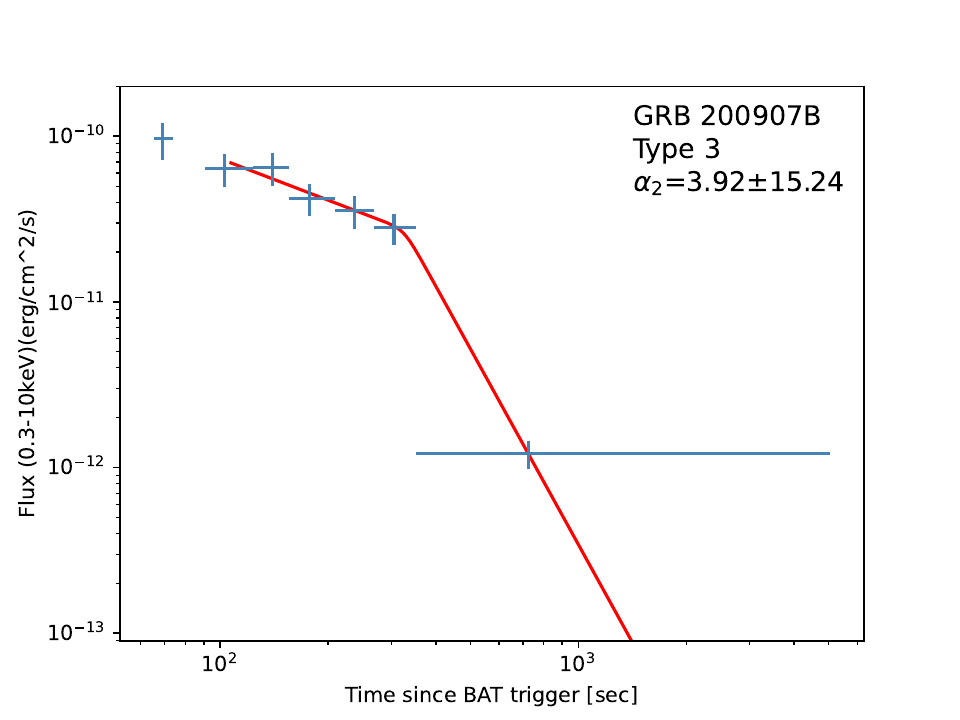}
  \end{minipage}\\

  \begin{minipage}[c]{0.33\textwidth}
     \centering
     \includegraphics[width=\textwidth, height=4.3cm]{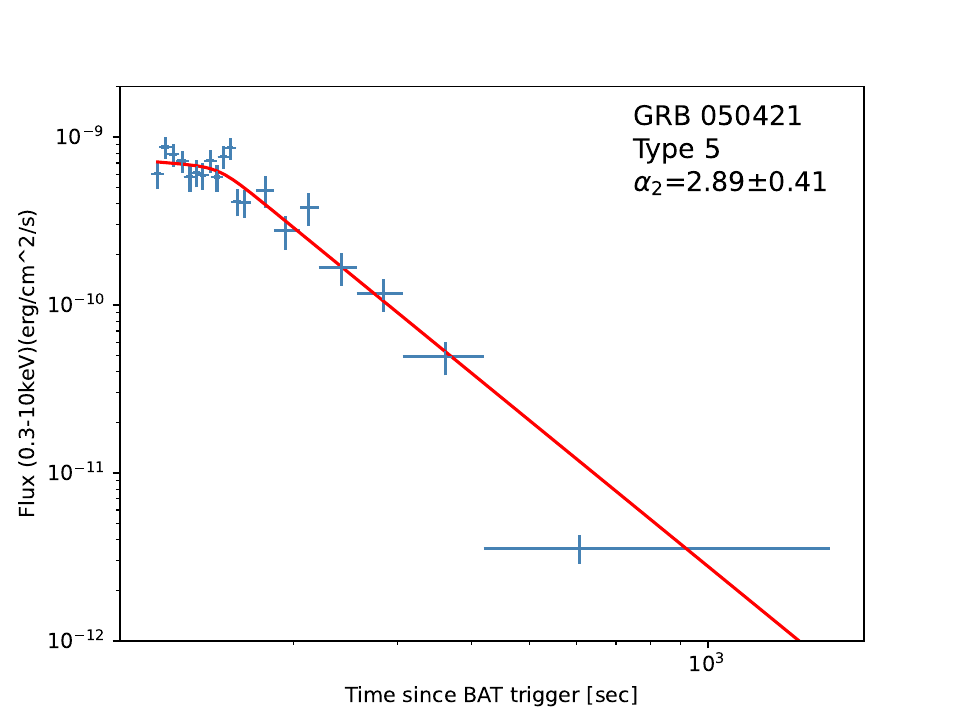}
  \end{minipage}
  \hfill
  \begin{minipage}[c]{0.33\textwidth}
     \centering      
     \includegraphics[width=\textwidth, height=4.3cm]{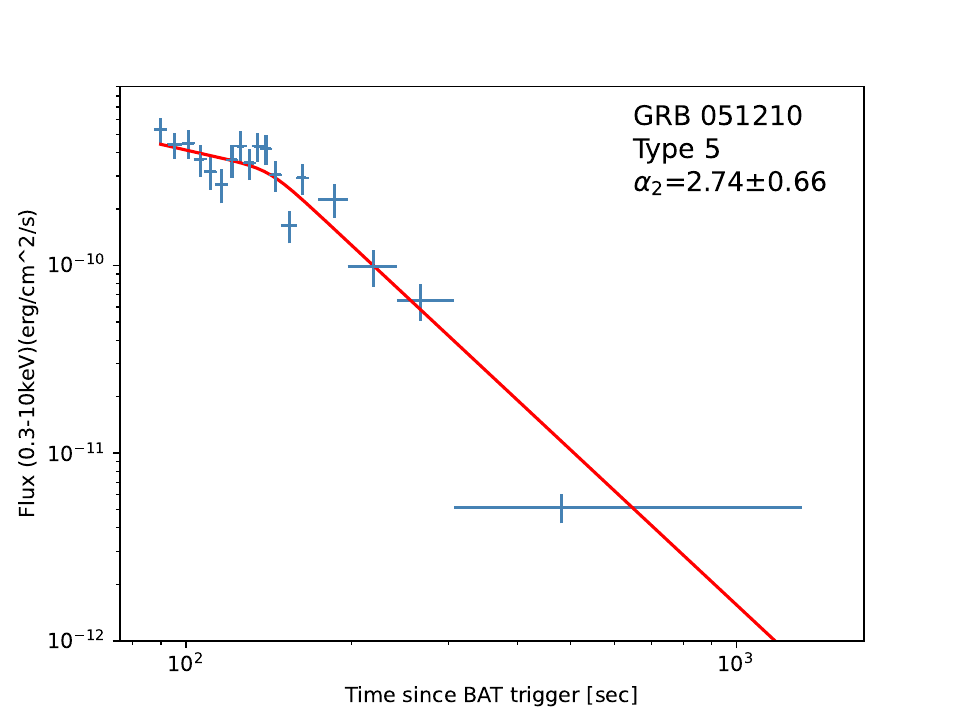}
  \end{minipage}
  \hfill
  \begin{minipage}[c]{0.33\textwidth}
     \centering      
     \includegraphics[width=\textwidth, height=4.3cm]{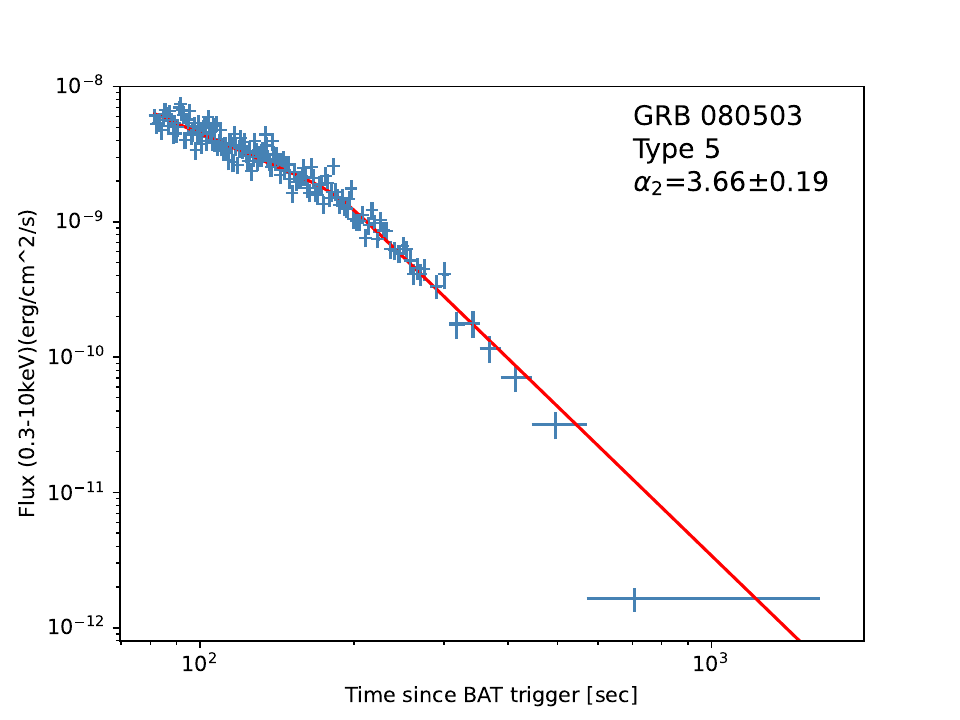}
  \end{minipage}\\

  \begin{minipage}[c]{0.33\textwidth}
     \centering
     \includegraphics[width=\textwidth, height=4.3cm]{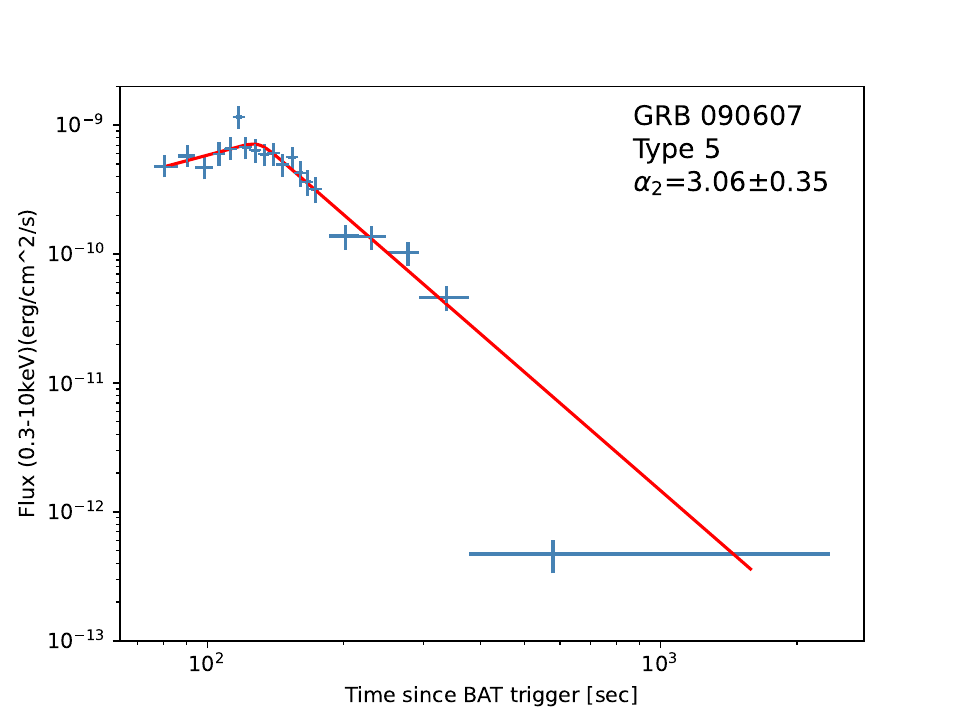}
  \end{minipage}
  \hfill
  \begin{minipage}[c]{0.33\textwidth}
     \centering      
     \includegraphics[width=\textwidth, height=4.3cm]{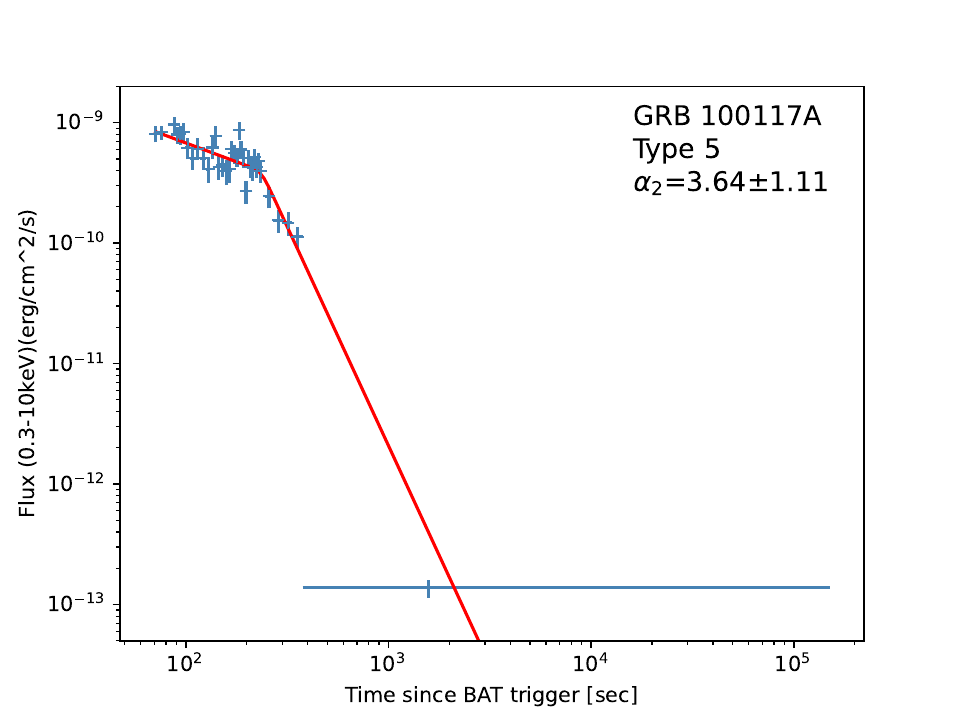}
  \end{minipage}
  \hfill
  \begin{minipage}[c]{0.33\textwidth}
     \centering      
     \includegraphics[width=\textwidth, height=4.3cm]{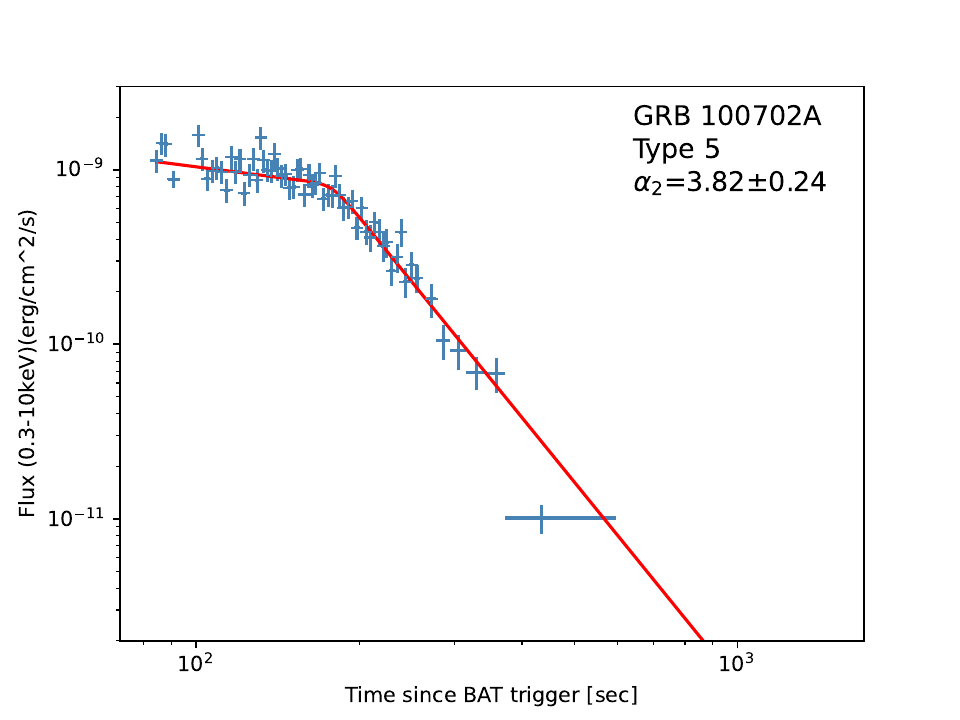}
  \end{minipage}\\

  \begin{minipage}[c]{0.33\textwidth}
      \centering
      \includegraphics[width=\textwidth, height=4.3cm]{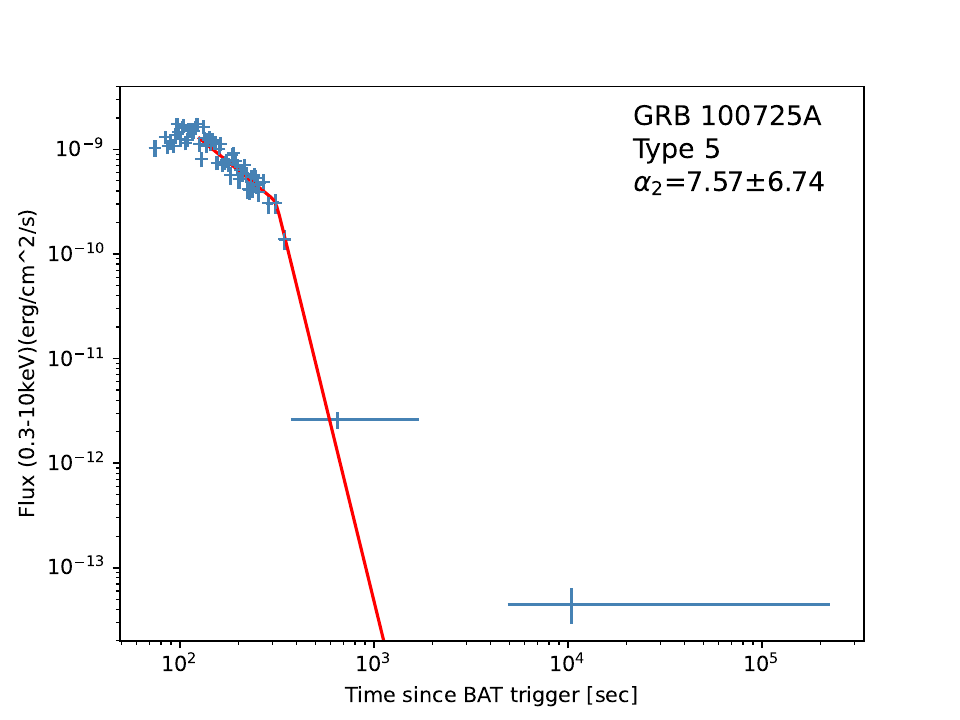}
   \end{minipage}
   \hfill
   \begin{minipage}[c]{0.33\textwidth}
      \centering      
      \includegraphics[width=\textwidth, height=4.3cm]{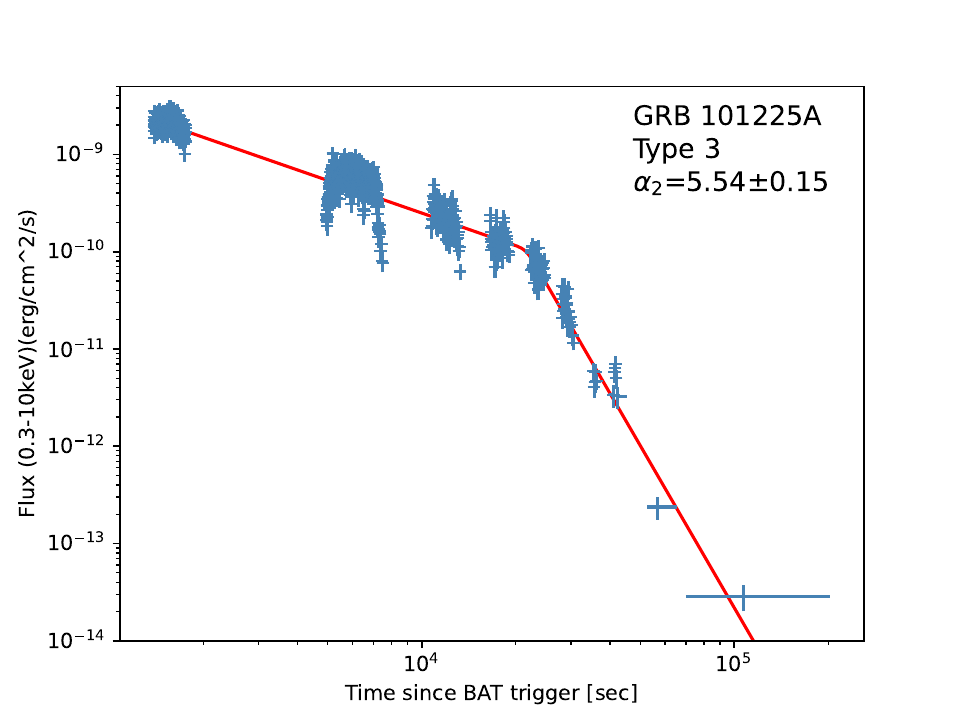}
   \end{minipage}
   \hfill
   \begin{minipage}[c]{0.33\textwidth}
      \centering      
      \includegraphics[width=\textwidth, height=4.3cm]{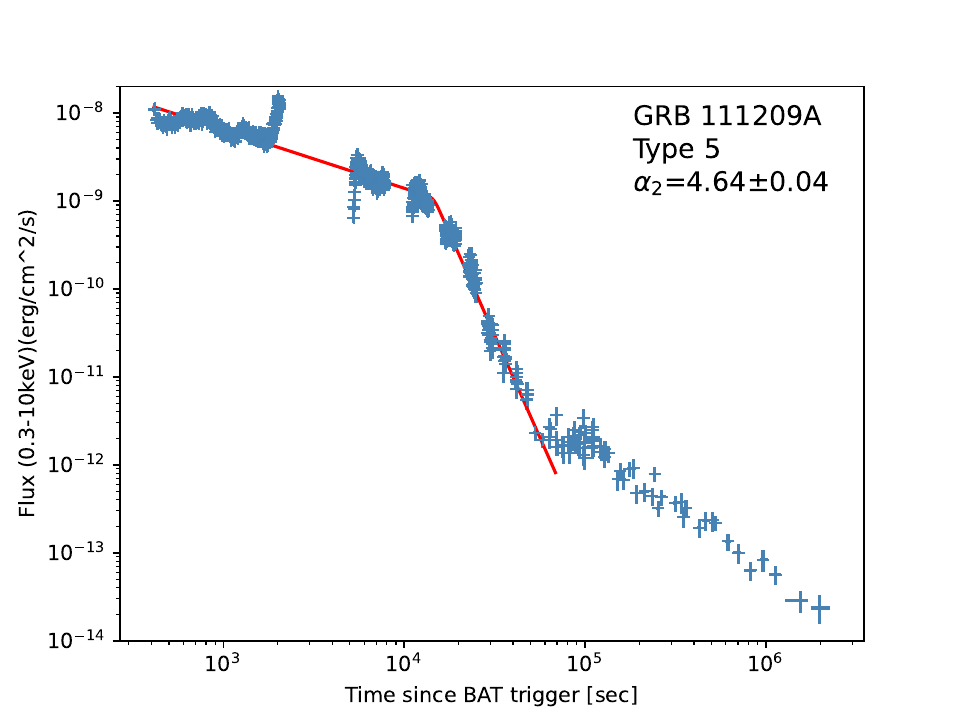}
   \end{minipage}\\
  
  \end{tabular}
  \caption{Continued: X-ray afterglow light curves.}
  \label{fig:lc_3}
\end{figure*}

\begin{figure*}[htbp]
  \addtocounter{figure}{-1}
  \begin{tabular}{ccc}

  \begin{minipage}[c]{0.33\textwidth}
     \centering
     \includegraphics[width=\textwidth, height=4.3cm]{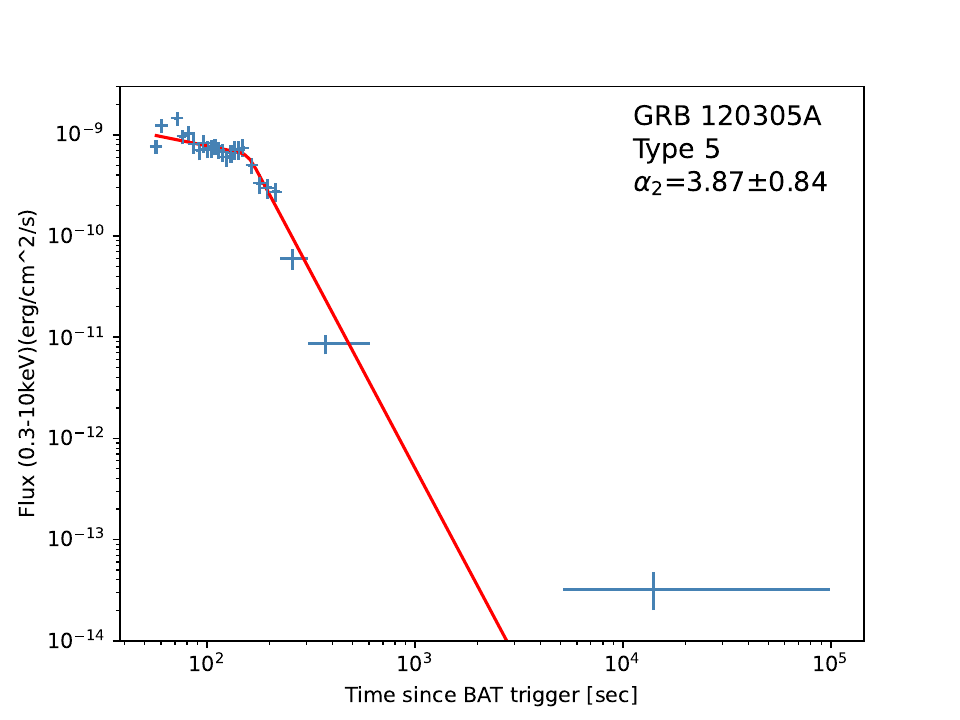}
  \end{minipage}
  \hfill
  \begin{minipage}[c]{0.33\textwidth}
     \centering      
     \includegraphics[width=\textwidth, height=4.3cm]{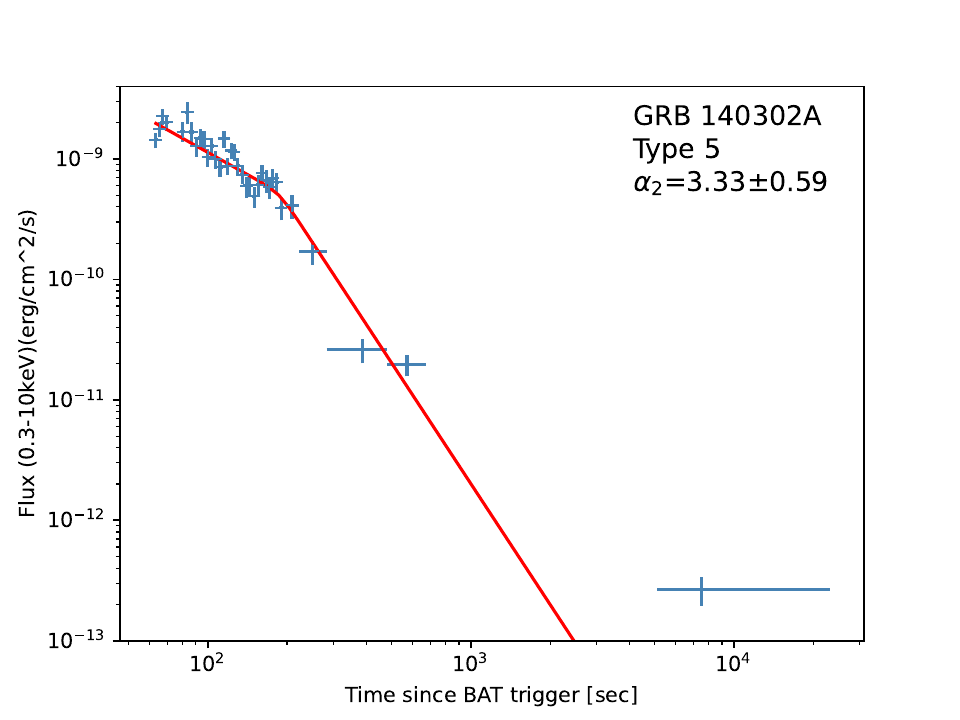}
  \end{minipage}
  \hfill
  \begin{minipage}[c]{0.33\textwidth}
     \centering      
     \includegraphics[width=\textwidth, height=4.3cm]{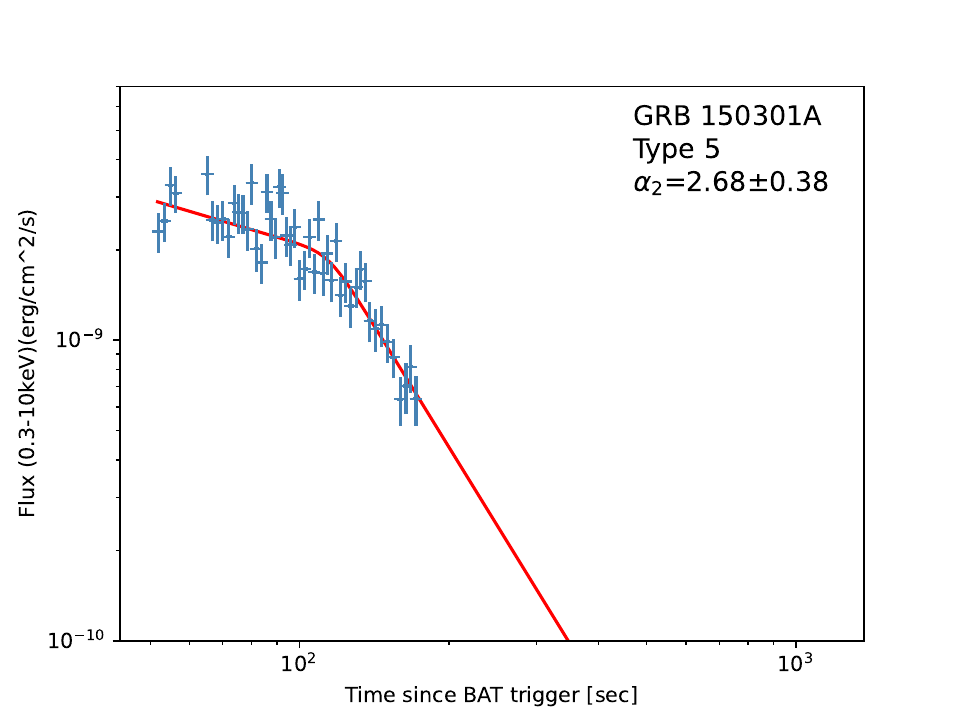}
  \end{minipage}\\
  
  \begin{minipage}[c]{0.33\textwidth}
     \centering
     \includegraphics[width=\textwidth, height=4.3cm]{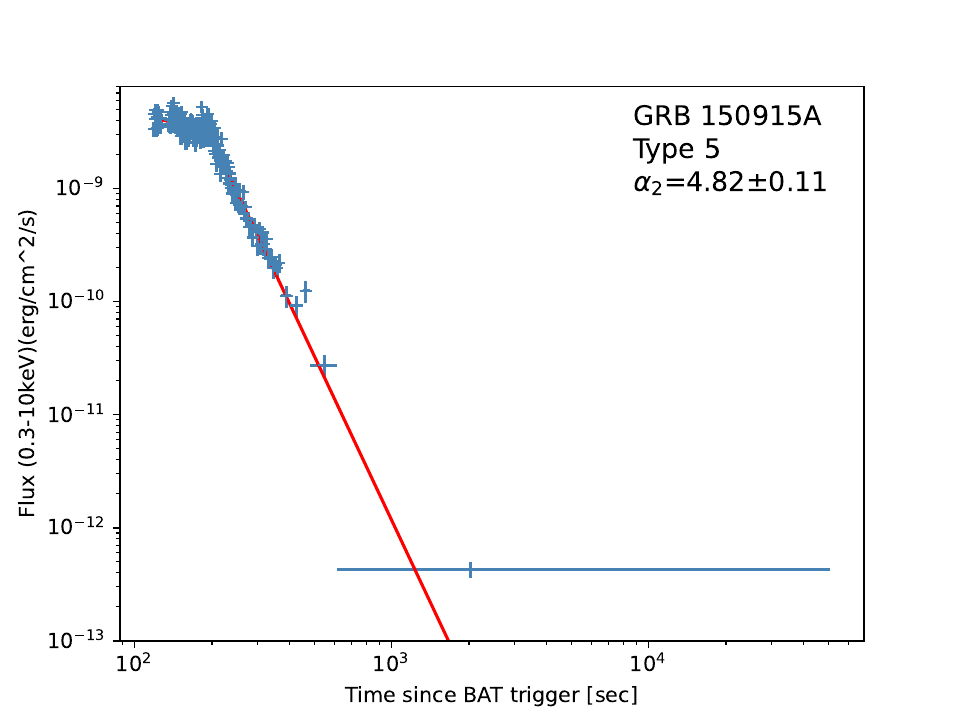}
  \end{minipage}
  \hfill
  \begin{minipage}[c]{0.33\textwidth}
     \centering      
     \includegraphics[width=\textwidth, height=4.3cm]{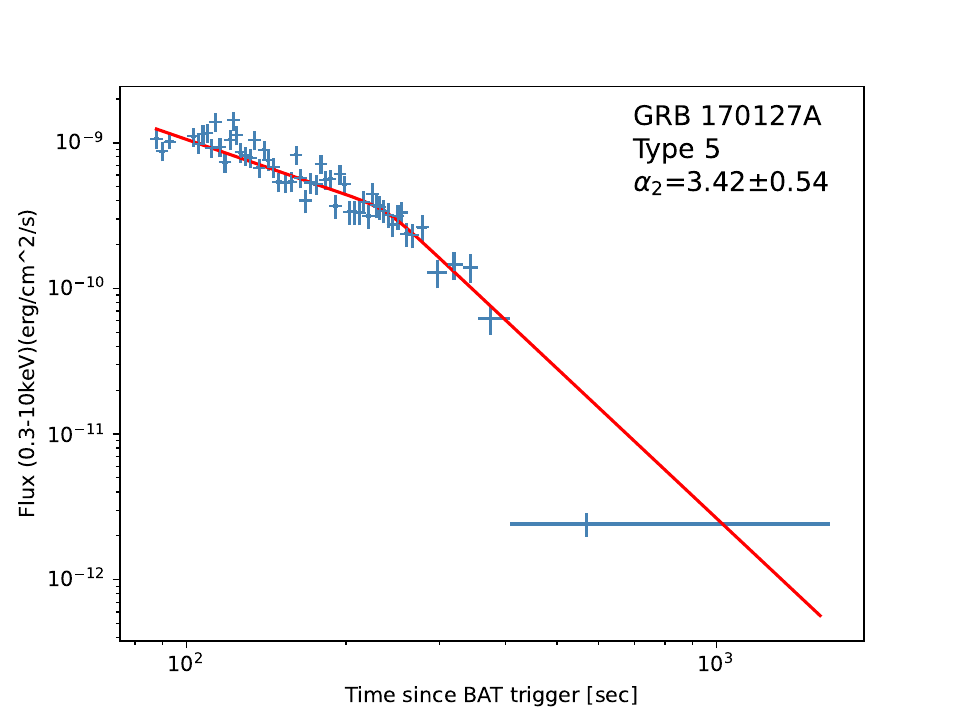}
  \end{minipage}
  \hfill
  \begin{minipage}[c]{0.33\textwidth}
     \centering      
     \includegraphics[width=\textwidth, height=4.3cm]{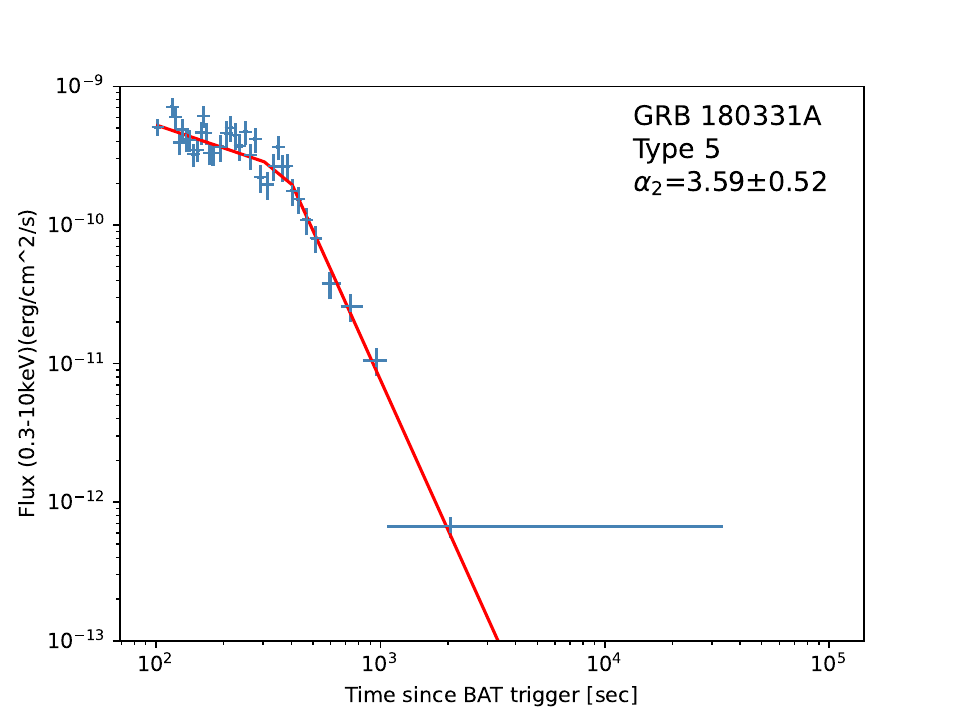}
  \end{minipage}\\

  \begin{minipage}[c]{0.33\textwidth}
     \centering
     \includegraphics[width=\textwidth, height=4.3cm]{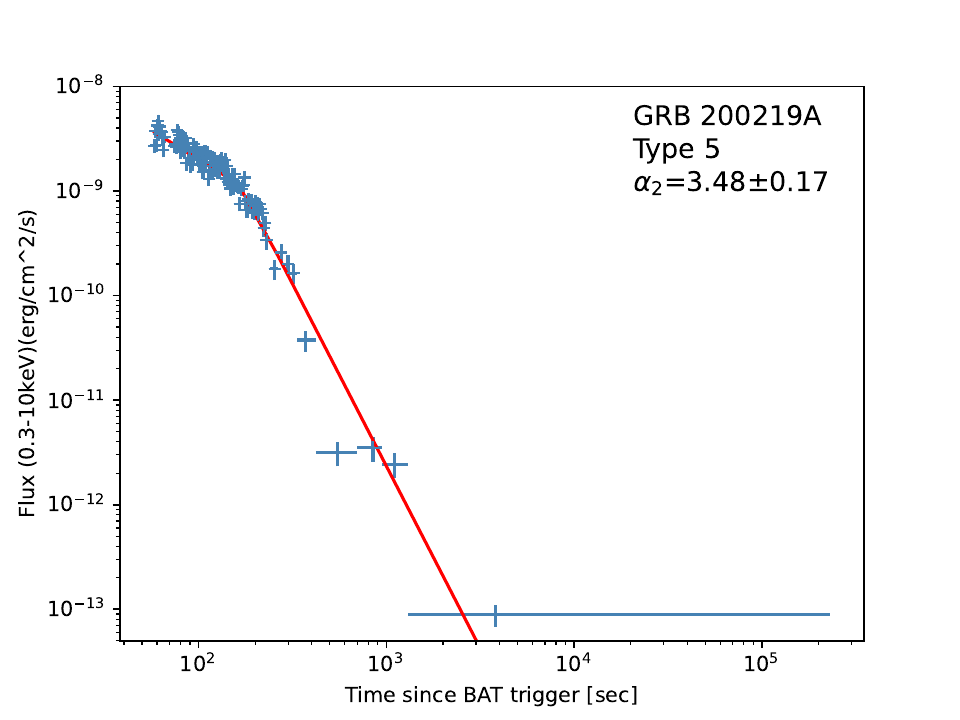}
  \end{minipage}
  \hfill
  \begin{minipage}[c]{0.33\textwidth}
     \centering      
     \includegraphics[width=\textwidth, height=4.3cm]{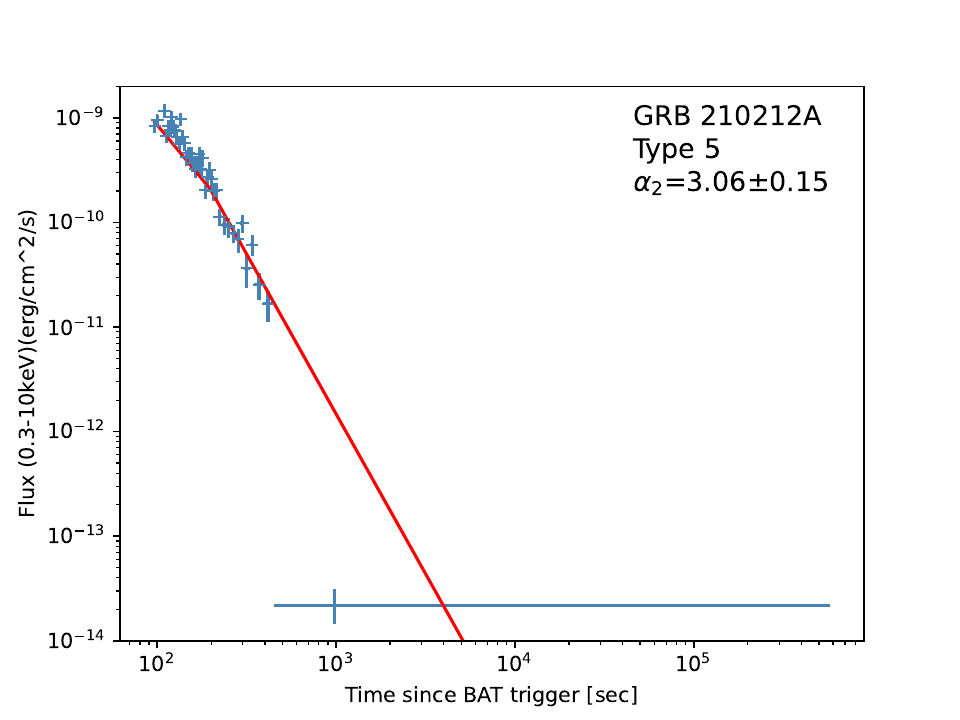}
  \end{minipage}
  \hfill
  \begin{minipage}[c]{0.33\textwidth}
     \centering      
     \includegraphics[width=\textwidth, height=4.3cm]{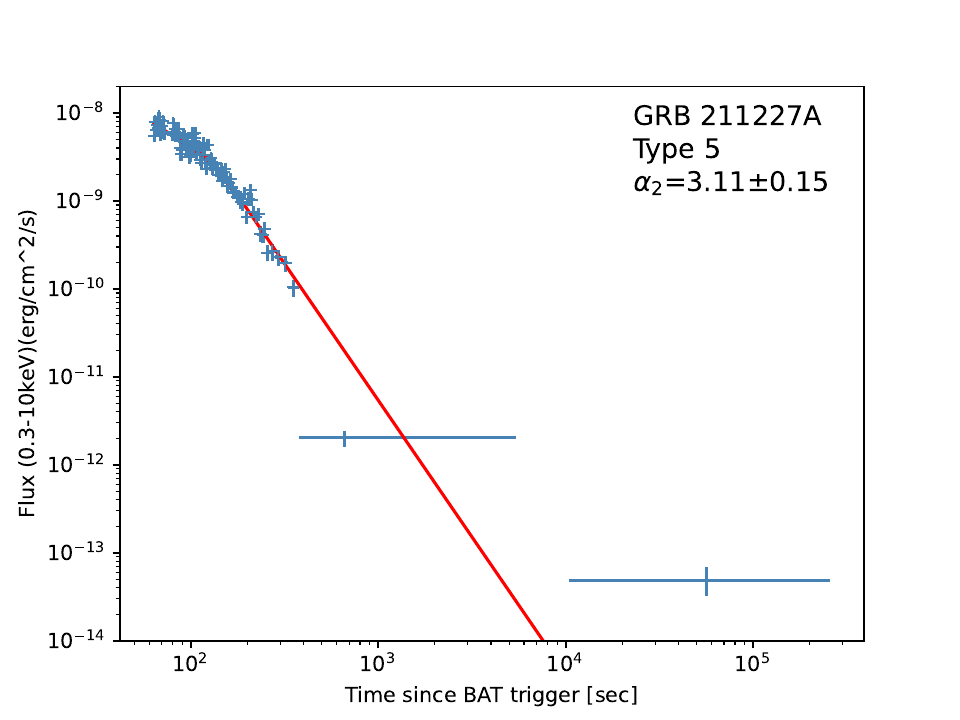}
  \end{minipage}\\

  \end{tabular}
  \caption{Continued: X-ray afterglow light curves.}
  \label{fig:lc_3}
\end{figure*}

\begin{table*}[htbp]
 \caption{\textit{Swift} observations and fitting result for our sample}
  \centering
  \small
   \begin{tabular}{lcccccc} \hline \hline
    GRB        & ${\alpha}_1$  & ${\alpha}_2$  & ${t_b}$[ks]     & ${{\chi}^2/dof}$  & RA            & Dec  \\ \hline
    \multicolumn{7}{c}{${\textbf{Type 1 Event}}$}\\
    060413     & 0.10 ${\pm}$ 0.06 & 2.92 ${\pm}$ 0.12 & 23.74 ${\pm}$ 0.75 & 127/103 & 19$^{\rm h}$ 25$^{\rm m}$ 07.8$^{\rm s}$& ${+}$13$^{\circ}$ 45$^{\prime}$ 30.5$^{\prime\prime}$\\ 
    060607A    & 0.45 ${\pm}$ 0.02 & 3.50 ${\pm}$ 0.10 & 12.75 ${\pm}$ 0.33 & 169/163 & 21$^{\rm h}$ 58$^{\rm m}$ 50.4$^{\rm s}$& ${-}$22$^{\circ}$ 29$^{\prime}$ 46.8$^{\prime\prime}$\\
    070110     & 0.30 ${\pm}$ 0.03 & 8.08 ${\pm}$ 0.57 & 20.62 ${\pm}$ 0.25 & 138/104 & 00$^{\rm h}$ 03$^{\rm m}$ 39.4$^{\rm s}$& ${-}$52$^{\circ}$ 58$^{\prime}$ 28.2$^{\prime\prime}$\\   
    070311     & ${-}$0.08 ${\pm}$ 0.35 & 3.07 ${\pm}$ 0.49 & 169.73 ${\pm}$ 30.52 & 8/9 & 05$^{\rm h}$ 50$^{\rm m}$ 08.27$^{\rm s}$& ${+}$03$^{\circ}$ 22$^{\prime}$ 29.6$^{\prime\prime}$\\   
    070429A    & 0.41 ${\pm}$ 0.08 & 2.70 ${\pm}$ 0.93 & 485.33 ${\pm}$ 78.54 & 30/14 & 19$^{\rm h}$ 50$^{\rm m}$ 49.0$^{\rm s}$& ${-}$32$^{\circ}$ 24$^{\prime}$ 18.0$^{\prime\prime}$\\
    071011     & ${-}$0.22 ${\pm}$ 0.52 & 3.27 ${\pm}$ 1.13 & 485.40 ${\pm}$ 45.18 & 8/9 & 00$^{\rm h}$ 33$^{\rm m}$ 33.1$^{\rm s}$& ${+}$61$^{\circ}$ 08$^{\prime}$ 01.2$^{\prime\prime}$\\
    080129     & ${-}$0.12 ${\pm}$ 0.34 & 3.19 ${\pm}$ 1.57 & 184.31 ${\pm}$ 25.66 & 5/3 & 07$^{\rm h}$ 01$^{\rm m}$ 08.2$^{\rm s}$& ${-}$07$^{\circ}$ 50$^{\prime}$ 46.5$^{\prime\prime}$\\
    081210     & 0.66 ${\pm}$ 0.05 & 2.28 ${\pm}$ 0.72 & 172.49 ${\pm}$ 35.08 & 15/21 & 04$^{\rm h}$ 41$^{\rm m}$ 56.1$^{\rm s}$& ${-}$11$^{\circ}$ 15$^{\prime}$ 26.0$^{\prime\prime}$\\
    090519     & 0.69 ${\pm}$ 0.24 & 2.38 ${\pm}$ 4.99 &  3.62 ${\pm}$  6.81 & 3/5 & 09$^{\rm h}$ 29$^{\rm m}$ 07.0$^{\rm s}$& ${+}$00$^{\circ}$ 10$^{\prime}$ 49.2$^{\prime\prime}$\\
    100219A    & 0.59 ${\pm}$ 0.05 & 3.10 ${\pm}$ 1.57 & 33.18 ${\pm}$  4.07 & 25/16 & 10$^{\rm h}$ 16$^{\rm m}$ 48.5$^{\rm s}$& ${-}$12$^{\circ}$ 34$^{\prime}$ 00.0$^{\prime\prime}$\\
    100902A    & 0.68 ${\pm}$ 0.03 & 3.58 ${\pm}$ 0.97 & 746.54 ${\pm}$ 93.77 & 32/37 & 03$^{\rm h}$ 14$^{\rm m}$ 31.0$^{\rm s}$& ${+}$30$^{\circ}$ 58$^{\prime}$ 43.6$^{\prime\prime}$\\
    110223A    & 0.60 ${\pm}$ 0.03 & 2.44 ${\pm}$ 1.32 & 538.88 ${\pm}$ 160.60 & 16/22 & 23$^{\rm h}$ 03$^{\rm m}$ 22.7$^{\rm s}$& ${+}$87$^{\circ}$ 33$^{\prime}$ 28.5$^{\prime\prime}$\\
    111022B    & 0.00 ${\pm}$ 0.23 & 2.68 ${\pm}$ 3.00 & 47.46 ${\pm}$ 12.96 &  1/1 & 07$^{\rm h}$ 15$^{\rm m}$ 51.5$^{\rm s}$& ${+}$49$^{\circ}$ 41$^{\prime}$ 00.9$^{\prime\prime}$\\
    120320A    & 0.02 ${\pm}$ 0.11 & 2.82 ${\pm}$ 0.91 & 74.58 ${\pm}$ 11.70 & 17/7  & 14$^{\rm h}$ 10$^{\rm m}$ 04.2$^{\rm s}$& ${+}$08$^{\circ}$ 41$^{\prime}$ 47.9$^{\prime\prime}$\\
    120521B    & 0.68 ${\pm}$ 0.05 & 2.55 ${\pm}$ 1.99 & 17.44 ${\pm}$  4.02 & 13/13 & 13$^{\rm h}$ 08$^{\rm m}$ 02.3$^{\rm s}$& ${-}$52$^{\circ}$ 45$^{\prime}$ 16.6$^{\prime\prime}$\\    
    120521C    & 0.46 ${\pm}$ 0.07 & 2.40 ${\pm}$ 2.96 & 21.29 ${\pm}$ 10.80 & 9/6 & 14$^{\rm h}$ 17$^{\rm m}$ 08.7$^{\rm s}$& ${+}$42$^{\circ}$ 08$^{\prime}$ 41.3$^{\prime\prime}$\\
    161011A    & 0.02 ${\pm}$ 0.04 & 3.12 ${\pm}$ 0.56 & 46.24 ${\pm}$  2.75 & 106/48 & 16$^{\rm h}$ 22$^{\rm m}$ 54.9$^{\rm s}$& ${+}$08$^{\circ}$ 18$^{\prime}$ 39.7$^{\prime\prime}$\\
    171004A    & 0.91 ${\pm}$ 0.23 & 3.00 ${\pm}$ 0.75 &  7.10 ${\pm}$  1.06 & 63/67 & 09$^{\rm h}$ 16$^{\rm m}$ 40.7$^{\rm s}$& ${+}$52$^{\circ}$ 41$^{\prime}$ 35.3$^{\prime\prime}$\\
    171209A    & 0.74 ${\pm}$ 0.07 & 2.71 ${\pm}$ 0.61 & 19.47 ${\pm}$  2.05 & 70/50 & 09$^{\rm h}$ 17$^{\rm m}$ 36.7$^{\rm s}$& ${-}$30$^{\circ}$ 31$^{\prime}$ 12.5$^{\prime\prime}$\\
    191031C    & 0.66 ${\pm}$ 0.05 & 4.12 ${\pm}$ 3.44 & 97.41 ${\pm}$ 50.48 & 33/18 & 07$^{\rm h}$ 43$^{\rm m}$ 28.8$^{\rm s}$& ${-}$62$^{\circ}$ 19$^{\prime}$ 31.1$^{\prime\prime}$\\
    201017A    & 0.28 ${\pm}$ 0.35 & 3.83 ${\pm}$ 15.14&  3.13 ${\pm}$  4.13 &  1/2  & 02$^{\rm h}$ 26$^{\rm m}$ 29.1$^{\rm s}$& ${+}$66$^{\circ}$ 40$^{\prime}$ 42.9$^{\prime\prime}$\\
    210323A    & 0.44 ${\pm}$ 0.06 & 3.49 ${\pm}$ 2.89 & 13.99 ${\pm}$ 10.67 & 12/11 & 21$^{\rm h}$ 11$^{\rm m}$ 47.2$^{\rm s}$& ${+}$25$^{\circ}$ 22$^{\prime}$ 09.1$^{\prime\prime}$\\ \hline   
    \multicolumn{7}{c}{${\textbf{Type 3 Event}}$}\\
    060801     & 0.84 ${\pm}$ 0.24 & 5.01 ${\pm}$ 4.24 &   0.34 ${\pm}$ 0.07   & 66/7      & 14$^{\rm h}$ 11$^{\rm m}$ 58.2$^{\rm s}$& ${+}$16$^{\circ}$ 59$^{\prime}$ 08.3$^{\prime\prime}$\\
    080919     & 0.98 ${\pm}$ 0.26 & 4.50 ${\pm}$ 5.85 &   0.37 ${\pm}$ 0.10   & 5/2       & 17$^{\rm h}$ 40$^{\rm m}$ 53.0$^{\rm s}$& ${-}$42$^{\circ}$ 22$^{\prime}$ 29.5$^{\prime\prime}$\\
    081029     & 0.45 ${\pm}$ 0.05 & 2.73 ${\pm}$ 0.20 &  17.96 ${\pm}$ 0.93   & 82/72     & 23$^{\rm h}$ 07$^{\rm m}$ 05.6$^{\rm s}$& ${-}$68$^{\circ}$ 09$^{\prime}$ 20.2$^{\prime\prime}$\\    
    090929B    & 0.96 ${\pm}$ 0.03 & 2.77 ${\pm}$ 0.79 & 242.58 ${\pm}$ 38.64  & 66/64     & 07$^{\rm h}$ 50$^{\rm m}$ 52.8$^{\rm s}$& ${-}$00$^{\circ}$ 39$^{\prime}$ 27.3$^{\prime\prime}$\\   
    160313A    & 1.87 ${\pm}$ 0.36 & 4.95 ${\pm}$ 4.62 & 0.31 ${\pm}$ 0.11 & 4/8          & 12$^{\rm h}$ 15$^{\rm m}$ 11.2$^{\rm s}$& ${+}$57$^{\circ}$ 17$^{\prime}$ 01.0$^{\prime\prime}$\\
    160623A    & 1.61 ${\pm}$ 0.08 & 2.99 ${\pm}$ 0.61 & 204.57 ${\pm}$ 41.66  & 72/81     & 21$^{\rm h}$ 01$^{\rm m}$ 11.54$^{\rm s}$& ${+}$42$^{\circ}$ 13$^{\prime}$ 15.5$^{\prime\prime}$\\
    160910A    & 1.21 ${\pm}$ 0.08 & 3.05 ${\pm}$ 1.97 & 493.68 ${\pm}$ 294.30 & 7/6       & 14$^{\rm h}$ 45$^{\rm m}$ 46.04$^{\rm s}$& ${+}$39$^{\circ}$ 04$^{\prime}$ 01.3$^{\prime\prime}$\\
    181023A    & 0.64 ${\pm}$ 0.02 & 3.04 ${\pm}$ 0.60 & 327.48 ${\pm}$ 25.79  & 146/52    & 16$^{\rm h}$ 33$^{\rm m}$ 17.4$^{\rm s}$& ${+}$19$^{\circ}$ 34$^{\prime}$ 51.2$^{\prime\prime}$\\    
    200306C    & 1.03 ${\pm}$ 0.03 & 2.49 ${\pm}$ 0.56 &   6.54 ${\pm}$ 1.52   & 68/46     & 13$^{\rm h}$ 14$^{\rm m}$ 18.6$^{\rm s}$& ${+}$11$^{\circ}$ 15$^{\prime}$ 47.7$^{\prime\prime}$\\
    200612A    & 1.06 ${\pm}$ 0.01 & 4.87 ${\pm}$ 2.30 &  27.04 ${\pm}$ 2.29   & 101/49    & 20$^{\rm h}$ 10$^{\rm m}$ 47.0$^{\rm s}$& ${-}$45$^{\circ}$ 21$^{\prime}$ 00.5$^{\prime\prime}$\\
    200907B    & 0.83 ${\pm}$ 0.42 & 3.92 ${\pm}$ 15.24 &  0.33 ${\pm}$ 0.26   & 1/2       & 05$^{\rm h}$ 56$^{\rm m}$ 04.5$^{\rm s}$& ${+}$06$^{\circ}$ 54$^{\prime}$ 17.7$^{\prime\prime}$\\ \hline 
    \multicolumn{7}{c}{${\textbf{Type 5 Event}}$}\\
    050421     & 0.29 ${\pm}$ 0.97 & 2.89 ${\pm}$ 0.41 & 0.15 ${\pm}$ 0.01 & 161/15       & 20$^{\rm h}$ 29$^{\rm m}$ 02.8$^{\rm s}$& ${+}$73$^{\circ}$ 39$^{\prime}$ 17.9$^{\prime\prime}$\\ 
    051210     & 0.66 ${\pm}$ 0.52 & 2.74 ${\pm}$ 0.66 & 0.14 ${\pm}$ 0.02 & 75/14        & 22$^{\rm h}$ 00$^{\rm m}$ 41.2$^{\rm s}$& ${-}$57$^{\circ}$ 36$^{\prime}$ 48.4$^{\prime\prime}$\\
    080503     & 1.63 ${\pm}$ 0.07 & 3.66 ${\pm}$ 0.19 & 0.18 ${\pm}$ 0.01 & 1185/146     & 19$^{\rm h}$ 06$^{\rm m}$ 28.6$^{\rm s}$& ${+}$68$^{\circ}$ 47$^{\prime}$ 35.4$^{\prime\prime}$\\   
    090607     & ${-}$0.95 ${\pm}$ 0.55 & 3.06 ${\pm}$ 0.35 & 0.13 ${\pm}$ 0.01 & 2793/16 & 12$^{\rm h}$ 44$^{\rm m}$ 40.2$^{\rm s}$& ${+}$44$^{\circ}$ 06$^{\prime}$ 20.8$^{\prime\prime}$\\
    100117A    & 0.62 ${\pm}$ 0.10 & 3.64 ${\pm}$ 1.11 & 0.24 ${\pm}$ 0.02 & 161/32       & 00$^{\rm h}$ 45$^{\rm m}$ 04.6$^{\rm s}$& ${-}$01$^{\circ}$ 35$^{\prime}$ 46.6$^{\prime\prime}$\\
    100702A    & 0.40 ${\pm}$ 0.12 & 3.82 ${\pm}$ 0.24 & 0.18 ${\pm}$ 0.00 & 86/59        & 16$^{\rm h}$ 22$^{\rm m}$ 47.2$^{\rm s}$& ${-}$56$^{\circ}$ 31$^{\prime}$ 53.1$^{\prime\prime}$\\
    100725A    & 1.53 ${\pm}$ 0.12 & 7.57 ${\pm}$ 6.74 & 0.31 ${\pm}$ 0.03 & 53/37        & 11$^{\rm h}$ 05$^{\rm m}$ 55.6$^{\rm s}$& ${-}$26$^{\circ}$ 40$^{\prime}$ 12.6$^{\prime\prime}$\\
    101225A    & 1.11 ${\pm}$ 0.01 & 5.54 ${\pm}$ 0.15 &  21.62 ${\pm}$ 0.22   & 4522/1250 & 00$^{\rm h}$ 00$^{\rm m}$ 47.5$^{\rm s}$& ${+}$44$^{\circ}$ 36$^{\prime}$ 01.2$^{\prime\prime}$\\ 
    111209A    & 0.67 ${\pm}$ 0.00 & 4.64 ${\pm}$ 0.04 & 14.56 ${\pm}$ 0.06 & 9625/1328   & 00$^{\rm h}$ 57$^{\rm m}$ 22.6$^{\rm s}$& ${-}$46$^{\circ}$ 48$^{\prime}$ 03.9$^{\prime\prime}$\\
    120305A    & 0.41 ${\pm}$ 0.15 & 3.87 ${\pm}$ 0.84 & 0.16 ${\pm}$ 0.01 & 22/20        & 03$^{\rm h}$ 10$^{\rm m}$ 08.7$^{\rm s}$& ${+}$28$^{\circ}$ 29$^{\prime}$ 31.0$^{\prime\prime}$\\
    140302A    & 1.22 ${\pm}$ 0.11 & 3.33 ${\pm}$ 0.59 & 0.19 ${\pm}$ 0.02 & 74/31        & 16$^{\rm h}$ 55$^{\rm m}$ 26.4$^{\rm s}$& ${-}$12$^{\circ}$ 52$^{\prime}$ 41.8$^{\prime\prime}$\\
    150301A    & 0.49 ${\pm}$ 0.13 & 2.68 ${\pm}$ 0.38 & 0.11 ${\pm}$ 0.01 & 54/44        & 16$^{\rm h}$ 17$^{\rm m}$ 13.1$^{\rm s}$& ${-}$48$^{\circ}$ 42$^{\prime}$ 47.2$^{\prime\prime}$\\    
    150915A    & 0.51 ${\pm}$ 0.11 & 4.83 ${\pm}$ 0.11 & 0.19 ${\pm}$ 0.00 & 240/191      & 21$^{\rm h}$ 18$^{\rm m}$ 37.9$^{\rm s}$& ${-}$34$^{\circ}$ 54$^{\prime}$ 48.1$^{\prime\prime}$\\    
    170127A    & 1.26 ${\pm}$ 0.09 & 3.42 ${\pm}$ 0.54 & 0.24 ${\pm}$ 0.02 & 1334/53      & 11$^{\rm h}$ 37$^{\rm m}$ 26.5$^{\rm s}$& ${-}$45$^{\circ}$ 50$^{\prime}$ 09.8$^{\prime\prime}$\\
    180331A    & 0.55 ${\pm}$ 0.11 & 3.59 ${\pm}$ 0.52 & 0.38 ${\pm}$ 0.03 & 37/31        & 04$^{\rm h}$ 24$^{\rm m}$ 05.9$^{\rm s}$& ${+}$13$^{\circ}$ 23$^{\prime}$ 56.1$^{\prime\prime}$\\
    200219A    & 1.07 ${\pm}$ 0.06 & 3.48 ${\pm}$ 0.17 & 0.17 ${\pm}$ 0.00 & 470/107      & 22$^{\rm h}$ 50$^{\rm m}$ 30.2$^{\rm s}$& ${-}$59$^{\circ}$ 07$^{\prime}$ 10.4$^{\prime\prime}$\\
    210212A    & 0.59 ${\pm}$ 0.61 & 3.06 ${\pm}$ 0.15 & 0.13 ${\pm}$ 0.01 & 48/35        & 04$^{\rm h}$ 49$^{\rm m}$ 04.0$^{\rm s}$& ${+}$07$^{\circ}$ 16$^{\prime}$ 10.9$^{\prime\prime}$\\
    211227A    & 1.39 ${\pm}$ 0.10 & 3.11 ${\pm}$ 0.15 & 0.14 ${\pm}$ 0.01 & 1924/97      & 08$^{\rm h}$ 48$^{\rm m}$ 35.7$^{\rm s}$& ${-}$02$^{\circ}$ 44$^{\prime}$ 07.0$^{\prime\prime}$\\ \hline
   \end{tabular}
  \label{table1}
\end{table*}

\newpage
\section{Discusion}
\begin{figure*}[htbp]
  \begin{center}
    \includegraphics[width=7.2cm,clip]{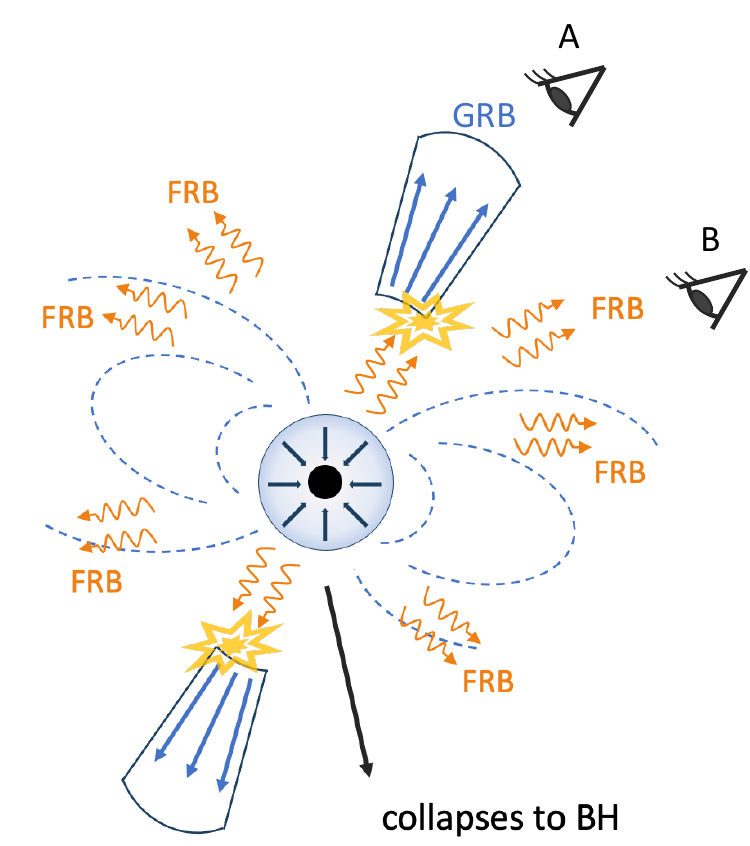}
  \end{center}
  \caption{A schematic picture of our model. GRBs could be detected from observer A. However, FRBs are not detectable since the GRB ejecta would absorb radio signals. On the other hand, FRBs could be observed from the direction of observer B.}
  \label{fig:frb_grb_gw}
\end{figure*}

The observer's direction could be a critical point in inquiring about the reason for the absence of FRB-associated GRBs. Generally, a GRB is observed when we see the jet from the on-axis direction (observer A of figure \ref{fig:frb_grb_gw}). Even if an FRB occurred, the GRB ejecta would absorb radio emission from the central engine. In contrast, it may be able to see an FRB without a GRB signal when an observer sees the event from an off-axis direction (observer B of figure \ref{fig:frb_grb_gw}). In this off-axis scenario with a BNS origin, if the event happens at a near distance, a GW signal could also be observed. This can be the case of FRB~20190425A (2.5 hrs after GW~190425) reported by \cite{Moroianu2023}. In our scenario, although it is difficult to detect a GRB with an FRB signal, the origin of FRBs could be the same as that of short GRBs, which is a BNS merger.

The environment of FRBs and GRBs is also a point to compare.
\cite{LiY2020} showed that most of the FRB host galaxies' stellar mass and star formation rate prefer a medium to old population, which implies that the environment of FRB is inconsistent with that of long GRBs but more consistent with short GRBs. On the other hand, the event rate of FRB ($R_{\rm FRB}$($L$ $>$ $10^{37}$ erg s$^{-1}$) ${\sim}$ $10^{7}-10^{8}$ Gpc$^{-3}$ yr$^{-1}$ \citep{Luo2020}) is much higher than that of long GRBs ($R_{\rm l-GRB}$ ${\sim}$ 1.3 Gpc$^{-3}$ yr$^{-1}$ \citep{Wanderman2010}) and short GRBs ($R_{\rm s-GRB}$ ${\sim}$ 7.5 Gpc$^{-3}$ yr$^{-1}$ \citep{ZhangGQ2018}). In our picture, if the open angle of the GRB jet was ${\theta}_j$ ${\sim}$ 5 degrees, the chance we could detect FRBs from the off-jet angle would be almost three orders of magnitude higher than the chance we observe an on-axis GRB. Therefore, only some FRBs could be explained by our scenario. However, our search of the {\it Swift} GRB samples is incomplete, considering the limited field of view and the sensitivity of the {\it Swift} Burst Alert Telescope (BAT) \citep{Barthelmy2005}. The peak energy flux of our samples observed by BAT ranged from 2.4 $\times 10^{-8}$ to 5.8 $\times 10^{-6}$ erg s$^{-1}$ cm$^{-2}$. We need to increase the samples by combining data from multiple GRB observatories and also search for weak GRBs by the upcoming new X-ray transient facility, such as the Einstein Probe \citep{Yuan2022} and the HiZ-GUNDAM mission \citep{Yonetoku2022}.

Non-repeating FRBs are the high-priority target for unraveling the connection between FRBs, GRBs, and GW events. 
However, it is difficult to catch these transient events because we never know where and when they will come. Also, the error range of the gravitational wave detector is as extensive as ${\sim}$100 deg${^2}$, and the detectors of GRB or GW sometimes look at a different sky than the telescopes of the FRB do. For this reason, the radio observatory with a large field of view is essential. 
Bustling Universe Radio Survey Telescope in Taiwan (BURSTT), a new fisheye radio software telescope with a large field of view of ${\sim10^4}$ deg${^2}$, can detect and localize ${\sim}$100 nearby FRBs per year \citep{Lin_BURSTT}. Thanks to the wide field of view, BURSTT could discover a large sample of FRBs and achieve immediate multi-wavelength and multi-messenger follow-up observation.

\section{Summary}
We investigate the case suggested by \cite{BingZhang2014} using the extensive X-ray afterglow data of {\it Swift}. 
We elasticated and selected 51 samples from the \textit{Swift}/XRT X-ray afterglow data. We found no GRB-associated FRBs in our samples.
In future work, we would like to combine data from multiple GRB observatories and compare the onset between GRBs and FRBs with a broader time window.
Also, we would like to apply a multivariate adaptive regression splines (MARS) technique \citep{Friedman1991} to improve our light curve fitting.
A radio telescope with a large field of view, such as BURSTT, and an upcoming high-sensitivity X-ray transient facility, such as the Einstein Probe and the HiZ-GUNDAM, are needed to unveil the association between FRB, GRB, and GW.

We thank the referee for their careful reading and their suggestions that substantially improved the quality of this paper.
We would like to thank T. Hashimoto and S. Yamasaki for variable comments. This research was supported by JST SPRING, Grant Number JPMJSP2103 (HS) and partially supported by JSPS KAKENHI Grant Nos. 22KJ2643 (YS).
This work made use of data supplied by the UK Swift Science Data Centre at the University of Leicester.


\end{document}